\documentclass[journal]{IEEEtran}
\IEEEoverridecommandlockouts
\usepackage{cite}
\usepackage{graphicx}
\usepackage{amsmath}
\usepackage{amssymb}
\usepackage{algorithmicx}
\usepackage{algorithm}
\usepackage{color}

\newtheorem{proposition}{Proposition}
\newtheorem{remark}{Remark}

\allowdisplaybreaks

\begin{document}

\title{MIMO Detection for Reconfigurable Intelligent Surface-Assisted Millimeter Wave Systems}

\author{\IEEEauthorblockN{Xi Yang, Chao-Kai Wen, and Shi Jin}}

\maketitle

\begin{abstract}

Millimeter wave (mmWave) band, or high frequencies such as THz, has large undeveloped band of spectrum. However, wireless channels over the mmWave band usually have one or two paths only due to the severe attenuation. The channel property restricts its development in the multiple-input multiple-output (MIMO) system, which can improve throughput by increasing the spectral efficiency. Recent development in reconfigurable intelligent surface (RIS) provides new opportunities to mmWave communications. In this study, we propose a mmWave system, which used low-precision analog-to-digital converters (ADCs), with the aid of several RIS arrays. Moreover, each RIS array has many reflectors with discrete phase shift. By employing the linear spatial processing, these arrays form a synthetic channel with increased spatial diversity and power gain, which can support MIMO transmission. We develop a MIMO detector according to the characteristics of the synthetic channel. RIS arrays can provide spatial diversity to support MIMO transmission, however, different number, antenna configuration, and deployment of RIS arrays affect the bit error rate (BER) performance. We present state evolution (SE) equations to evaluate the BER of the proposed MIMO detector in the different cases. The BER performance of indoor system is studied extensively through leveraging by the SE equations. We reveal numerous insights about the RIS effects and discuss the appropriate system settings. In addition, our results demonstrate that the low-cost hardware, such as the 3-bit ADCs of the receiver side and the 2-bit uniform discrete phase shift of the RIS arrays, only moderately degenerate the system performance.

\end{abstract}


\section{Introduction}

\begin{figure}
\centering
\includegraphics[scale = 0.6]{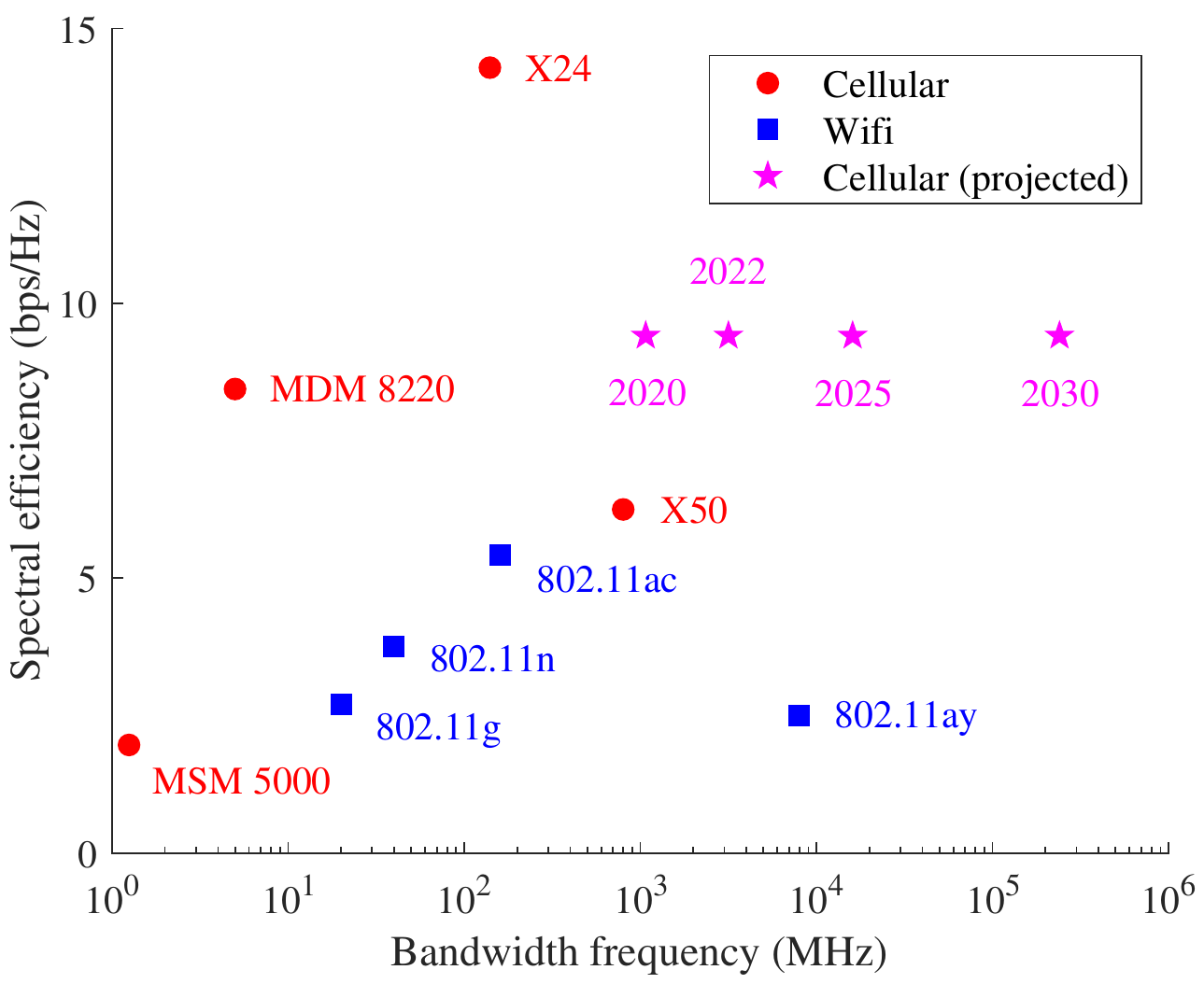}
\caption{Evolution of spectral efficiencies and bandwidths with cellular systems and Wifi.}
\label{bandwidth_se}
\end{figure}

According to the Ericsson outlook \cite{Ericsson}, mobile data traffic presents a $30$ percent compound annual growth rate between 2018 and 2024. The traffic growth increases in average data usage per smartphone from $5.6$ gigabytes (GB) to $22.5$ GB per month. Reference \cite{bandwidth} investigates the historical evolution trends on the basis of cellular systems and WiFi in the past 25 years and reveals that the peak data rate with cellular systems is predicted to reach $151.1$ Gbps and $2259.9$ Gbps by 2025 and 2030, respectively. Fig.~\ref{bandwidth_se} illustrates the cellular data rate with Qualcomm modems \cite{Qualcomm}, WiFi data rate with IEEE 802.11 standards \cite{Wifi}, and the projected data rate of cellular systems in the near future \cite{bandwidth}. The required frequency bandwidth of future wireless communication system grows rapidly along with time under the fixed spectral efficiency of 5G-NR air-link specifications. In 2030, the cellular system is predicted to need a bandwidth of 240.22 GHz to satisfy the data rate demand.

With the growing demand for data rate, future wireless communication systems are expected to exploit the undeveloped spectrum in high frequencies, such as millimeter wave (mmWave) \cite{MMW_1,MMW_2,MMW_3} and THz regimes. Although the large bandwidth guarantees a high data rate, promising mmWave communication also faces many challenges. One of the problems is severe attenuation in the high frequency. Compared with the transmissions over the sub-6G band, mmWave communication suffers from high attenuation, especially that caused by low-reflective surfaces \cite{MMW_4}. The wireless channel over mmWave band usually has one or two paths only. Given the limited spectral efficiency caused by poor spatial diversity, mmWave communication requires a large bandwidth to realize high data rate. However, obtaining such a large frequency band to meet the rapidly growing demands of future wireless communications is very difficult, even in high frequencies.

Fortunately, recent research on reconfigurable intelligent surface (RIS) provides a brand new approach to improve the performance of wireless communication systems by changing the electromagnetic propagation environment rather than adapting to it. RIS is a passive and power-saving artificial material that includes digitally controlled reprogrammable reflector arrays \cite{RIS_1,RIS_2,RIS_3}. The most recent RIS can work in high frequencies, even in the THz spectrum \cite{RIS_4}. RIS can improve the propagation environments by manipulating the incident electromagnetic wave, thereby providing huge gains and new opportunities to wireless communications \cite{RIS_5,RIS_6,PL}. Several studies have investigated the potential of data transmission for wireless communication system with the aid of RIS \cite{Meta_1,Meta_2,Meta_3,Meta_4}. Results show that the RIS can offer huge gains to the traditional wireless communication system, resulting in enhanced energy efficiency and reduced number of active antennas at the base station (BS). Researchers also investigated the potential uses of RIS as transmitter \cite{Meta_5} and receiver \cite{Meta_6}. Reference \cite{Meta_7} designed a practical communication system, where RIS is used as the wireless transmitter; the proposed radio frequency (RF) chain-free transmitter realizes 6.144 Mbps data rate over the air in the experiments. According to \cite{Meta_5,Meta_6,Meta_7}, RIS-based transceiver reduce the signal processing complexity, hardware cost, and power consumption. The hardware imperfections of RIS-assisted communication system are also studied in \cite{Meta_3,Meta_8,Meta_9,Meta_10,Meta_11}. These works reveal the potential huge gains provided by the RIS. However, studies about the RIS-assisted mmWave system are limited. Communication over the mmWave band needs specific transmission scheme designs to fulfill the potentials of RIS because of the special channel characteristics. As mentioned above, we have a particular interest in the potential spatial diversity gains provided by the RIS. We are curious about how to obtain the spatial diversity gains with the aid of RIS and how to design the appropriate detector for a RIS-assisted mmWave system.

Another problem of the mmWave communication system is expensive hardware. The receiver needs analog-to-digital converter (ADC) with high sampling rate because of the large bandwidth used. However, the power consumption of ADC increases quadratically with sampling rate at above 100 MHz \cite{ADC_1}. High-precision ADC further aggravate the situation. The fabrication cost and power consumption of ADC increase exponentially with the number of quantization bits \cite{ADC_2}. Many studies have investigated the wireless communication systems with low-precision ADCs through rate analysis \cite{EE_1,EE_2,EE_3,EE_4}, channel estimation \cite{CE_1,CE_2,CE_3}, and data detection \cite{JCD,DD_1,DD_2}. Low-precision ADCs only moderately reduce the system performance compared with the full-precision ones. Hence, investigating RIS-assisted mmWave system with low-precision ADCs is reasonable than costly and power-hungry high-precision ADCs. RIS with continuous phase shift reflectors has many benefits, however, it may cause expensive cost on the hardware. Therefore, we also interested in the potential implementation of RIS with discrete phase shift reflectors.

Research on mmWave systems with increased spectral efficiency and reduced hardware cost is very important for future wireless communications. In this study, we propose a point-to-point RIS-assisted mmWave system with the aid of several RIS arrays. By employing the linear spatial processing for each RIS, these arrays form a synthetic channel with enhanced spatial diversity and power gain and can support the multiple-input multiple-output (MIMO) transmission. Based on the characteristics of the synthetic channel, we investigate the MIMO detection of the proposed RIS-assisted mmWave system. Furthermore, we reveal the connections between the bit error rate (BER) performance and the synthetic channel, as determined by the deployment of RIS arrays. Our main contributions are as follows:

\begin{itemize}

\item We consider a coded MIMO transmission scheme to deal with possibly low spatial diversity of the synthetic channel and satisfy the power consumption demand of user equipment (UE). We derive a MIMO detector based on the framework of Bayesian inference to obtain an effective detection algorithm for the coded system. Furthermore, we present state evolution (SE) equations with descent and ascent processes as an effective tool to determine the BER performance of the MIMO detector.

\item We study the RIS effects of an indoor system, where the RIS arrays are deployed around a circle on the surface of ceiling. A large number of RIS arrays offer enhanced spatial diversity and received power, which remarkably improve the performance of the system. Furthermore, we find that the uniform linear array (ULA) case has better performance than the uniform rectangular array (URA) case because of limited diversity of elevation. However, ULA case is relatively sensitive to the orientations of the arrays of UE and BS. We show that the proposed deployment of RIS arrays is robust for different UE locations since the RIS arrays are spatially separated. Finally, we discuss the effect of the direct path to the proposed system.

\item We investigate the potential implementation of low-cost hardware, which includes the low-precision ADCs of BS and the discrete phase shift of RIS arrays. We find that the 3-bit ADCs of the BS and the 2-bit uniform phase shift of the RIS arrays only moderately increases the BER compared with the full-precision ADCs and continuous phase shift. Hence, the proposed RIS-assisted mmWave system can be deployed with low-cost hardware, since the cheap components do not cause large performance degradation.

\end{itemize}

\emph{Notations}: Capital and lowercase boldface letters denote matrices and vectors, respectively. For any matrix ${\bf A}$, ${{\bf A}^H}$ denotes the conjugate transpose of ${\bf A}$ and ${\rm tr}{\left({\bf A}\right)}$ represents its trace. ${{\bf I}}$ is the identity matrix, ${{\bf 0}}$ represents the zero vector, ${\rm E}{\left\{{\cdot}\right\}}$ denotes the expectation operator, and ${\Phi}{\left( {\cdot} \right)}$ is the cumulative Gaussian distribution function. Moreover, $\left\langle {\bf{v}} \right\rangle$ computes the average value of vector ${\bf v}$, and ${\rm d}{\left\{ {\bf Q} \right\}}$ calculates the average value of diagonal elements of matrix ${\bf Q}$. Besides, ${\mathcal{N}}{\left( {x; \mu, \sigma^2} \right)}$ is the real Gaussian distribution function with mean $\mu$ and variance $\sigma^2$.

\section{System Model}

\begin{figure*}
\centering
\includegraphics[scale = 0.5]{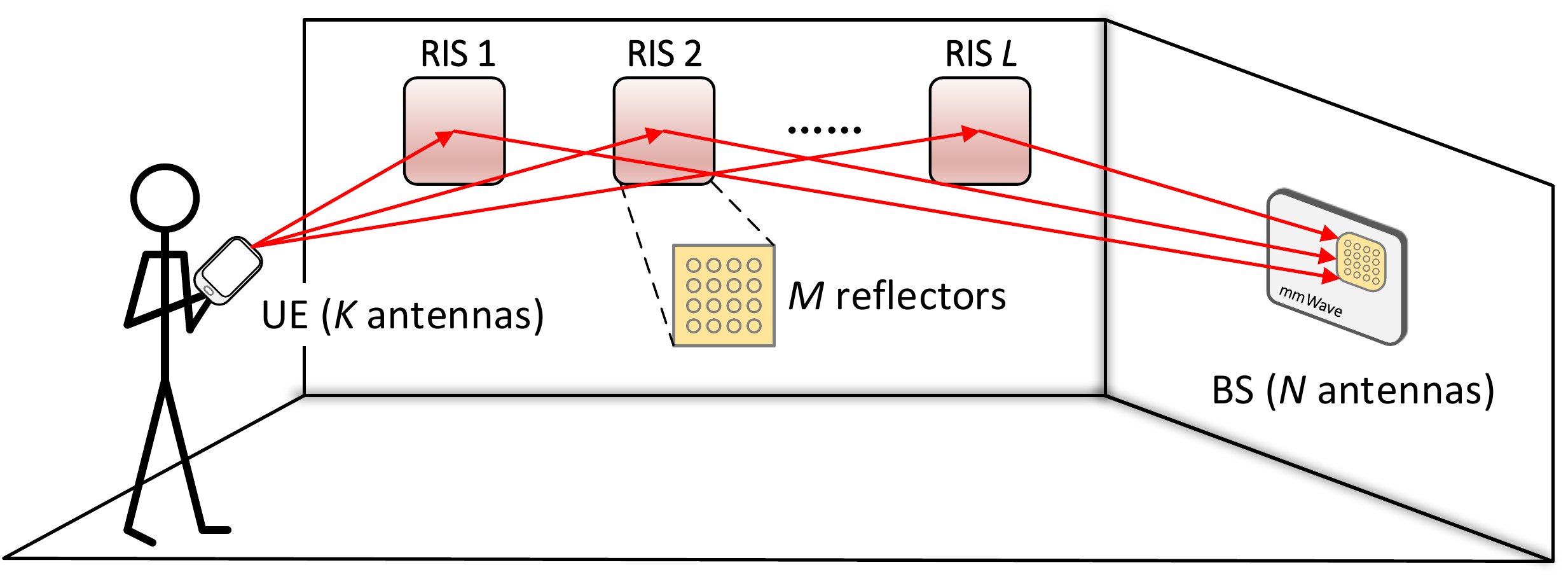}
\caption{Model of the RIS-assisted mmWave system.}
\label{system_model}
\vspace{-0.2cm}
\end{figure*}

Fig.~\ref{system_model} shows a point-to-point mmWave system where a $K$-antenna UE transmits the signal to an $N$-antenna BS with the aid of several RIS arrays, where each RIS has $M$ reflectors. Moreover, BS uses ${\rm B}$-bit ADC at each RF chain. RIS arrays are deployed in a distributed manner to perform spatial processing in order to improve the electromagnetic propagation environments and provide additional spatial diversity. Among various spatial processing strategies, a simple but easily deployed linear spatial processing technology is used. In particular, each RIS array performs equal gain combining (EGC) to enhance the received signal power at the BS. The wireless channel between the UE and BS via $i$th RIS array can be written as
\begin{equation}\label{RIS_channel}
{\bf A}_i = {\frac{1}{\sqrt K}}{\beta_i}{{\bf g}_i{\bf a}_i^H}{\bf \Omega}_i{{\bf b}_i{\bf h}_i^H},
\end{equation}
where $\beta_i$ and ${\bf \Omega}_i$ denote the path loss factor and the diagonal phase shift matrix of $i$th RIS array, respectively. We consider the line-of-sight (LoS) transmission between the UE and BS via the RIS.\footnote{For simplicity, we use the channel model with a single LoS path only. The general channel model with multiple paths can be used as long as angle-of-arrival (AoA) and angle-of-depature (AoD) of main path can be obtained. In the following analysis, it shows that the EGC spatial processing only enhance the power of main path. Therefore, we can ignore other reflection paths.} Assuming that the antenna arrays of UE, RIS, and BS are ULA, steering vectors are given as
\begin{subequations}
\begin{align}
{\bf g}_i & = {\left[ {1, e^{j2{\pi}{\frac{d_{\rm bs}}{\lambda}}{\sin}{\theta}_{\rm bs}^i}, {\cdots}, e^{j2{\pi}{\frac{d_{\rm bs}}{\lambda}}(N{-}1){\sin}{\theta}_{\rm bs}^i}} \right]^H}, \\
{\bf a}_i & = {\left[ {1, e^{j2{\pi}{\frac{d_{\rm ris}}{\lambda}}{\sin}{\phi}_{\rm ris}^i}, {\cdots}, e^{j2{\pi}{\frac{d_{\rm ris}}{\lambda}}(M{-}1){\sin}{\phi}_{\rm ris}^i}} \right]^H}, \\
{\bf b}_i & = {\left[ {1, e^{j2{\pi}{\frac{d_{\rm ris}}{\lambda}}{\sin}{\theta}_{\rm ris}^i}, {\cdots}, e^{j2{\pi}{\frac{d_{\rm ris}}{\lambda}}(M{-}1){\sin}{\theta}_{\rm ris}^i}} \right]^H}, \\
{\bf h}_i & = {\left[ {1, e^{j2{\pi}{\frac{d_{\rm ue}}{\lambda}}{\sin}{\phi}_{\rm ue}^i}, {\cdots}, e^{j2{\pi}{\frac{d_{\rm ue}}{\lambda}}(K{-}1){\sin}{\phi}_{\rm ue}^i}} \right]^H},
\end{align}
\end{subequations}
where ${\theta}_{\rm bs}^i$, ${\phi}_{\rm ris}^i$, ${\theta}_{\rm ris}^i$, and ${\phi}_{\rm ue}^i$ are AoA of BS, AoD of RIS, AoA of RIS, and AoD of UE, respectively, and $d_{\rm ue}$, $d_{\rm ris}$, and $d_{\rm bs}$ are antenna spacing of UE, RIS, and BS, respectively. Besides, $\lambda$ is the wavelength.

\begin{table*}
\centering
\caption{Singular values of different channels (one trial).}
\begin{tabular}{|l|c|c|c|c|c|c|c|c|}
\hline
& \multicolumn{8}{|c|}{ Singular Values }     \\ \hline
Rayleigh Channel & 2.110 & 1.833 & 1.589 & 1.452 & 1.245 & 0.997 & 0.819 & 0.584 \\ \hline
Synthetic Channel & 2.521 & 1.757 & 1.714 & 1.678 & 0.649 & 0.616 & 0.038 & 0.015 \\ \hline 
\end{tabular}
\end{table*}

By estimating the wireless channels\footnote{Reference \cite{Katabi} realized an indoor mmWave system with the aid of a mirror to support the mobile virtual reality headset. The proposed method in \cite{Katabi} can be used in the mmWave system of our work. The channel between the UE and the RIS and the channel between the RIS and the BS are fully characterized by the locations of the UE, RIS, and BS. Given that the locations of the RIS and the BS are fixed, the channel between the RIS and the BS is known. The UE transmits the pilot to the BS via the RIS, and the BS can estimate the channel between the UE and the RIS. When we estimate the channel for a specific RIS array, we can turn off other RIS arrays to avoid interference. We can use the AoAs (azimuth and elevation) of two RIS arrays to position the location of the UE and use the location of the UE to infer the channels between the UE and other RIS arrays. Note that we set a wireless or wired controller for each RIS array to receive control messages from the BS. In this work, we assume the perfect channel state information is available.}, RIS can use EGC spatial processing
\begin{align}
{\bf \Omega}_i = & {\rm diag}{\left[ 1, e^{j2{\pi}{\frac{d_{\rm ris}}{\lambda}}({\sin}{\theta}_{\rm ris}^i{-}{\sin}{\phi}_{\rm ris}^i)}, \right.} \nonumber\\
& \ \ \ \ \ \ \ \ \ \ \ \ \ \ \ \ \ \ {\left. {\cdots}, e^{j2{\pi}{\frac{d_{\rm ris}}{\lambda}}(M{-}1)({\sin}{\theta}_{\rm ris}^i{-}{\sin}{\phi}_{\rm ris}^i)} \right].} \label{EGC_processing}
\end{align}
Plugging (\ref{EGC_processing}) into (\ref{RIS_channel}), we have
\begin{equation}\label{beamforming}
{\bf A}_i = {\frac{1}{\sqrt K}}M{\beta_i}{{\bf g}_i{\bf h}_i^H},
\end{equation}
which implies that the RIS contributes to enhancement of the received signal of a scale factor $M$ at the BS. Equation (\ref{beamforming}) shows the beamforming gain of $M$ of the RIS \cite{PL}. For convenience, we use the ULA case to illustrate the beamforming gain of the RIS. In practice, the RIS is usually designed as a URA. We can obtain the same beamforming gain for the URA case by employing a similar EGC spatial processing, such as (\ref{EGC_processing}), as indicated in Section~IV. In the proposed linear spatial processing scheme, the beamforming design is offloaded to the RIS. The EGC spatial processing provides considerable power gain, but the spatial diversity is rank one for a single RIS array. A single RIS only support single-input multiple-output transmission. We use a group of $L$ RIS arrays to realize MIMO transmission, and the wireless channel between the UE and BS is given as
\begin{equation}
{\bf A} = {\sum\limits_{i = 1}^L {\bf A}_i} = {\frac{1}{\sqrt K}}{\sum\limits_{i = 1}^L M{\beta_i}{\bf g}_i{\bf h}_i^H},
\end{equation}
where the $i$th RIS array performs its own EGC spatial processing.\footnote{In our work, we focus on single user transmission. The proposed mmWave system can be extended to a multiuser case. In fact, an RIS array can only provide a beamforming gain to a specific user or, more exactly, a specific spatial position (of a user). If two users are near each other, then the interference is high. We should assign different frequency spectra for different users, but each RIS array can serve two users. If two users are far from each other, then the interference is extremely small. Thus, we can assign the same frequency spectra for these users. However, each RIS array can only serve a specific user because the positions of users are different.} Since ${\rm B}$-bit ADCs are used, the received signal at the BS can be written as
\begin{equation}\label{GLM}
{\bf \tilde y} = {\sf Q_c}{\left( {\bf z} + {\bf w} \right)} = {\sf Q_c}{\left( {\bf A}{\bf x} + {\bf w} \right)}.
\end{equation}
${\bf x}$ is the transmit signal vector where each element is drawn from a set of normalized constellations, ${\bf w}$ denotes the complex Gaussian noise vector with the noise level $v_w$, and ${\sf Q_c}{\left( {\cdot} \right)}$ represents the ${\rm B}$-bit complex-valued quantizer. ${\sf Q_c}{\left( {\cdot} \right)}$ consists of two ${\rm B}$-bit real-valued quantizers for real and imaginary parts of the received signal. Each real-valued quantizer maps a real-valued input to one of the ${2^{\rm B}}$ bins, which are characterized by a set of ${2^{\rm B} - 1}$ thresholds ${\left[{{r_1}, {r_2}, {\cdots}, {r_{{2^{\rm B}}{-}1}}}\right]}$ such that ${-}{\infty} = {r_0} < {r_1} < {r_2} < {\cdots} < {r_{{2^{\rm B}}{-}1}} < {r_{{2^{\rm B}}}} = {\infty}$.

By exploiting the spatial diversity provided by several RIS arrays, we argue that the synthetic channel ${\bf A}$ can support the MIMO transmission. Compared with the Rayleigh channel under the rich-scattering propagation environments, the synthetic channel ${\bf A}$ has a {\em large beamforming gain}, which is increased quadratically with $M$, but {\em less spatial diversity}, especially when the number of RIS arrays ($L$) is small \cite{Spatial_diversity}. For example, Table 1 shows the singular values of Rayleigh and synthetic channels with normalized power, where $K = 8$, $L = 10$, and $N = 16$. The synthetic channel is not a rank-one matrix, which meets the basic requirement of the MIMO system, but has two weak singular values compared with the Rayleigh channel. Since we expect to use the RIS arrays as less as possible, the condition number of the synthetic channel will even be larger. Hence, MIMO transmission is possible under the synthetic channel, but the specific transmission scheme and detector are needed.

In this study, we use coding scheme to mitigate the less spatial diversity of synthetic channel and further reduce the transmit power of UE and derive the corresponding MIMO detector. We focus on two important issues of the proposed RIS-assisted mmWave system:

\begin{itemize}

\item[a)] {\em RIS effects:} The synthetic channel depends on the steering vectors of arrays. Hence, different orientations of arrays result in a different wireless channel. The spatial diversity of synthetic channel under the ULA case only depends on the diversity of azimuth, but that of the URA case depends on the diversities of azimuth and elevation. Different numbers of RIS arrays and UE locations cause a different synthetic channel. The direct path also has important effect to the synthetic channel. Therefore, we want to obtain a general analysis framework to determine the performance of the proposed MIMO detector under different environments to obtain effective system designs. In Section III, we present the analysis framework for the proposed MIMO detector, and the RIS effects are discussed in Section IV for an indoor system.

\item[b)] {\em ADC and Phase Quantization:} Although the low-precision ADC has low fabrication cost and power consumption, they can reduce the performance. To specify the tradeoff between the performance and the cost, we evaluate the performance of the proposed MIMO detector under different numbers of quantization bits and compare the performance of the quantized case with that of the unquantized one. Equation (\ref{EGC_processing}) implies that the perfect continuous phase shift is needed for each reflector. In practical application, phase shift will be interfered by phase noise, which has considerable negative effects on MIMO systems \cite{PN1,PN2}. Moreover, a continuous phase shift may cause high hardware cost. In the present study, we determine the impacts of perfect but discrete phase shift and investigate the potentials of deployment of low-cost RIS.

\end{itemize}

\section{MIMO Detection for RIS-Assisted mmWave System}

To mitigate the less spatial diversity of synthetic channel and decrease the power consumption of UE, we consider a coded transmission scheme with convolutional code. In a typical coding scheme for the convolutional code, the information bitstream ${\bf b}$ is transferred to the symbol stream ${\bf x}$ after encoding ($\bf s$), interleaving ($\bf d$), and modulating ($\bf x$). Furthermore, the symbol stream ${\bf x}$ is transmitted over several channels within a coherence time block, where the synthetic channel remains constant. In this section, we first propose a low-complexity iterative decoding algorithm based on the expectation consistent (EC) inference \cite{EC}, which takes the form of expectation propagation. Next, we compare the proposed algorithm with the related works and explain why we use the coding scheme. Finally, we present the SE equations as an analytical tool to evaluate the performance of the proposed algorithm.

\subsection{Detection Algorithm}

\begin{figure*}
\centering
\resizebox{5.5in}{!}
{
\begin{tabular}{ccc}
\includegraphics*{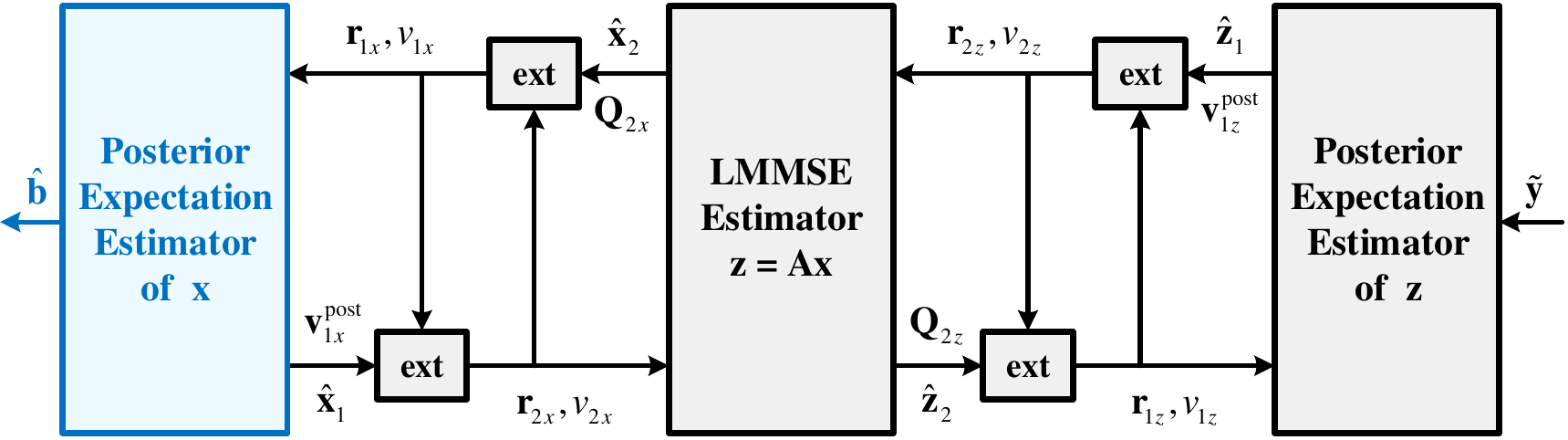} \\
\large (a) \\
\includegraphics*{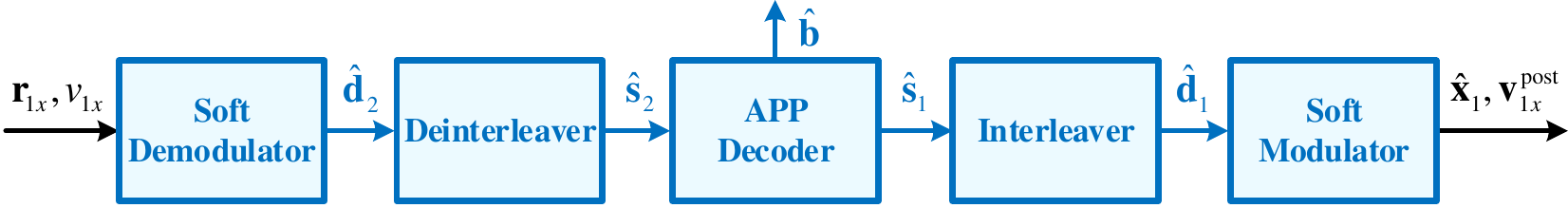} \\
\large (b) \\
\end{tabular}
}
\caption{(a) Block diagram of the GEC-C algorithm. (b) Details of the posterior expectation estimator of ${\bf x}$.}
\label{block_diagram_1_2}
\end{figure*}

In this subsection, we propose a MIMO detector, which is called generalized expectation consistent with coding (GEC-C), to recover the transmit bitstream. Fig.~\ref{block_diagram_1_2}(a) shows the block diagram of GEC-C, and Fig.~\ref{block_diagram_1_2}(b) details the estimate of ${\bf x}$. The main idea of the GEC-C is to perform low-complexity iterative decoding based on the {\em local minimum mean square error (MMSE) estimate} and the {\em turbo-type interference cancellation}.

Fig.~\ref{block_diagram_1_2}(a) shows that GEC-C has three estimators, where each of them performs the local MMSE estimate, and the extrinsic information is propagated based on the turbo-type interference cancellation. In the reverse link, GEC-C solves the quantization problem and obtains the MMSE estimate by using the posterior expectation estimator of ${\bf z}$ at first. The linear MMSE (LMMSE) estimator is applied to obtain the MMSE estimate of ${\bf x}$. Finally, the extrinsic information of ${\bf x}$ is mapped to the log-likelihood ratio (LLR) of encoded bit stream ${\bf s}$ and send to the a posteriori probability (APP) decoder after soft demodulating and deinterleaving. In the forward link, the inverse operations are taken. At first, the APP decoder performs the BCJR algorithm \cite{BCJR}, and the LLRs of information bitstream ${\bf b}$ and encoded bit stream ${\bf s}$ are computed. The MMSE estimate of ${\bf x}$ is calculated using the interleaver and the soft modulator. Finally, the LMMSE estimator is used to obtain the MMSE estimate of ${\bf z}$.

We use the turbo-type interference cancellation to avoid correlations among the iterations. The extrinsic information of one link is obtained by subtracting the counterpart of another link from the MMSE estimate. After several iterations, the algorithm will be converged, and the estimate of bitstream ${\bf b}$ can be obtained by using the hard decision for the APP decoder output ${\bf \hat b}$ (LLR). The procedures of GEC-C are specified in Algorithm~1. In the remaining part, we specify the computations of each estimator of GEC-C.

\begin{algorithm}
\begin{footnotesize}
\caption{GEC-C algorithm for MIMO detection.}
{\bf Input:} Trellis of convolutional code, synthetic channel matrix ${\bf A}$, and the conditional pdf of quantization ${p{\left({{\left.{\bf \tilde y}\right|}{\bf z}}\right)}}$.\\
{\bf Initial:} Extrinsic information ${{\bf r}_{2x}} = {\bf 0}$, ${v_{2x}} = 1$, ${{\bf r}_{1z}} = {\bf 0}$, ${v_{1z}} = {\frac{1}{N}}{\rm tr}{\left( {{\bf A}{\bf A}^H} \right)}$.\\
{\bf while} {$t < T_{\max}$} {\bf do}
\begin{enumerate}
\item Compute the posterior mean and variance from the quantization
\begin{equation}
{\bf \hat z}_1 = {\rm E}{\left\{ {\left. {\bf z} \right| {\bf r}_{1z}{,}v_{1z}} \right\}} \ \ {\rm and} \ \ {\bf v}_{1z}^{\rm post} = {\rm Var}{\left\{ {\left. {\bf z} \right| {\bf r}_{1z}{,}v_{1z}} \right\}},
\end{equation}
Compute the extrinsic information of ${\bf z}$
\begin{subequations}
\begin{align}
v_{2z} & = {\frac{1}{\frac{1}{\langle {{\bf v}_{1z}^{\rm post}} \rangle} - {\frac{1}{v_{1z}}}}}, \label{ext_z_mean}\\
{\bf r}_{2z} & = v_{2z}{\left( {\frac{{\bf \hat z}_1}{\langle {{\bf v}_{1z}^{\rm post}} \rangle} - {\frac{{\bf r}_{1z}}{v_{1z}}}} \right)}, \label{ext_z_var}
\end{align}
\end{subequations}
\item Compute the posterior mean and variance from the linear transform
\begin{subequations}
\begin{align}
{\bf Q}_{2x} & = {\left( {\frac{\bf I}{v_{2x}} + \frac{{\bf A}^H{\bf A}}{v_{2z}}} \right)}^{-1}, \label{lt_m_r}\\
{\bf \hat x}_2 & = {\bf Q}_{2x}{\left( {\frac{{\bf r}_{2x}}{v_{2x}} + \frac{{\bf A}^H{\bf r}_{2z}}{v_{2z}}} \right)}, \label{lt_v_r}
\end{align}
\end{subequations}
Compute the extrinsic information of ${\bf x}$
\begin{equation}
v_{1x} = {\frac{1}{\frac{1}{{\rm d}{\left\{ {\bf Q}_{2x} \right\}}} - {\frac{1}{v_{2x}}}}} \ \ {\rm and} \ \ {\bf r}_{1x} = v_{1x}{\left( {\frac{{\bf \hat x}_2}{{\rm d}{\left\{ {\bf Q}_{2x} \right\}}} - {\frac{{\bf r}_{2x}}{v_{2x}}}} \right)},
\end{equation}
\item Compute the LLRs from the code constraint
\begin{align}
{\bf \hat d}_2 = {\rm SDM}{\left( {{\bf r}_{1x}, v_{1x}} \right)} \ \to \ {\bf \hat s}_2 = {\rm DIL}{\left( {{\bf \hat d}_2} \right)} \nonumber\\
\to \ {\left\{ {\bf \hat b}, {\bf \hat s}_1 \right\}} = {\rm APP}{\left( {{\bf \hat s}_2} \right)} \ \to \ {\bf \hat d}_1 = {\rm IL}{\left( {{\bf \hat s}_1} \right)},
\end{align}
Compute the posterior mean and variance from the LLR
\begin{align}
{\bf \hat x}_1 = {\rm E}{\left\{ {\left. {\bf x} \right| {\bf \hat d}_1} \right\}} \ \ {\rm and} \ \ {\bf v}_{1x}^{\rm post} = {\rm Var}{\left\{ {\left. {\bf x} \right| {\bf \hat d}_1} \right\}},
\end{align}
Compute the extrinsic information of ${\bf x}$
\begin{subequations}
\begin{align}
v_{2x} & = {\frac{1}{\frac{1}{\langle {{\bf v}_{1x}^{\rm post}} \rangle} - {\frac{1}{v_{1x}}}}}, \label{ext_m}\\
{\bf r}_{2x} & = v_{2x}{\left( {\frac{{\bf \hat x}_1}{\langle {{\bf v}_{1x}^{\rm post}} \rangle} - {\frac{{\bf r}_{1x}}{v_{1x}}}} \right)}, \label{ext_v}
\end{align}
\end{subequations}
\item Compute the posterior mean and variance from the linear transform
\begin{subequations}
\begin{align}
{\bf Q}_{2z} & = {\bf A}{\bf Q}_{2x}{\bf A}^H, \label{lt_m_f}\\
{\bf \hat z}_2 & = {\bf A}{\bf \hat x}_2, \label{lt_v_f}
\end{align}
\end{subequations}
Compute the extrinsic information of ${\bf z}$
\begin{equation}
v_{1z} = {\frac{1}{\frac{1}{{\rm d}{\left\{ {\bf Q}_{2z} \right\}}} - {\frac{1}{v_{2z}}}}} \ \ {\rm and} \ \ {\bf r}_{1z} = v_{1z}{\left( {\frac{{\bf \hat z}_2}{{\rm d}{\left\{ {\bf Q}_{2z} \right\}}} - {\frac{{\bf r}_{2z}}{v_{2z}}}} \right)}.
\end{equation}
\end{enumerate}
{\bf Output:} Hard decision of the LLR stream ${\bf \hat b}$.
\end{footnotesize}
\end{algorithm}

\subsubsection{Posterior Expectation Estimator of ${\bf z}$}

GEC-C uses the expectation propagation method to develop an iterative decoding procedure. The posterior expectation estimator of ${\bf z}$ uses the extrinsic information (${\bf r}_{1z}$ and $v_{1z}$) from the LMMSE estimator to construct the following posterior probability density function (pdf)
\begin{equation}\label{posterior_pdf_z}
p{\left( {{\left. {\bf z} \right|}{{\bf r}_{1z}}, {v_{1z}}} \right)} = \frac{{p{\left( {\left. {\bf \tilde y} \right|}{\bf z} \right)}{\exp}{\left( { - \frac{{\left\| {\bf z} - {{\bf r}_{1z}} \right\|}^2}{v_{1z}}} \right)}}}{{{\int} {p{\left( {\left. {\bf \tilde y} \right|}{\bf z} \right)}{\exp}{\left( { - \frac{{\left\| {\bf z} - {{\bf r}_{1z}} \right\|}^2}{v_{1z}}} \right)}{d{\bf z}}} }}.
\end{equation}
The posterior mean ${\bf \hat z}_1$ and variance ${\bf v}_{1z}^{\rm post}$ are taken over the pdf (\ref{posterior_pdf_z}). ${\bf \hat z}_1$ and ${\bf v}_{1z}^{\rm post}$ are computed elementwisely because the conditional pdf ${p{\left({{\left.{\bf \tilde y}\right|}{\bf z}}\right)}}$ is separable. Additional details are presented in Appendix A of \cite{JCD}. In the GEC-C, the output extrinsic information (\ref{ext_z_mean}) and (\ref{ext_z_var}), which is obtained by subtracting the input extrinsic information from the posterior mean and variance, is acted as an interference cancellator like those in the belief propagation type iterative decoding algorithms. However, the belief propagation based iterative decoding algorithms perform the interference cancellation on the posterior pdf directly rather than the mean and variance with respect to the posterior pdf.

\subsubsection{LMMSE Estimator of Linear Transform}

The LMMSE estimator has the input extrinsic information from the posterior expectation estimators of ${\bf x}$ and ${\bf z}$. The posterior pdf of the LMMSE estimator is therefore given by
\begin{align}
& p{\left( {{\left. {\bf x} \right|}{{\bf r}_{2x}}, {v_{2x}}, {{\bf r}_{2z}}, {v_{2z}}} \right)} \nonumber\\
= & \frac{{\exp}{\left( {-}\frac{{\left\| {\bf x}{-}{{\bf r}_{2x}} \right\|}^2}{v_{2x}} \right)}{\exp}{\left( {-}\frac{{\left\| {\bf Ax}{-}{{\bf r}_{2z}} \right\|}^2}{v_{2z}} \right)}}{{\int}{{\exp}{\left( {-}\frac{{\left\| {\bf x}{-}{{\bf r}_{2x}} \right\|}^2}{v_{2x}} \right)}{\exp}{\left( {-}\frac{{\left\| {\bf Ax}{-}{{\bf r}_{2z}} \right\|}^2}{v_{2z}} \right)}}d{\bf x}}.
\end{align}
The posterior mean and variance of $\bf x$ can be obtained as (\ref{lt_m_r}) and (\ref{lt_v_r}) by employing some algebraic operations. Moreover, those of ${\bf z}$ can be obtained as (\ref{lt_m_f}) and (\ref{lt_v_f}) in the linear space ${\bf z} = {\bf Ax}$ (one-channel use). The complexity issue of matrix inversion in (\ref{lt_m_r}), (\ref{lt_v_r}), (\ref{lt_m_f}), and (\ref{lt_v_f}) can be alleviated using the one-time SVD trick ${\bf A} = {\bf U}{\bf S}{\bf V}^H$ \cite{VAMP}. Notably, we abuse the extrinsic information (${\bf r}_{2x}$ and ${\bf r}_{2z}$) in (\ref{lt_m_r}), (\ref{lt_v_r}), (\ref{lt_m_f}), and (\ref{lt_v_f}) for only one channel use, and complete transmission spans various channels.

\subsubsection{Posterior Expectation Estimator of ${\bf x}$}

Similarly, the posterior pdf of the posterior expectation estimator of ${\bf x}$ is given as
\begin{equation}\label{pdf_coded}
p{\left( {{\left. {\bf x} \right|}{{\bf r}_{1x}}{,}{v_{1x}}} \right)} = \frac{{p{\left( {\left. {\bf x} \right|} {\rm trellis} \right)}{\exp} {\left( {-}{\frac{{\left\| {{\bf x}{-}{\bf r}_{1x}} \right\|}^2}{v_{1x}}} \right)}}}{{\int}{p{\left( {\left. {\bf x} \right|} {\rm trellis} \right)}{\exp}{\left( {-}\frac{{\left\| {{\bf x}{-}{\bf r}_{1x}} \right\|}^2}{v_{1x}} \right)}d{\bf x}}},
\end{equation}
where the prior distribution $p{\left( {\left. {\bf x} \right|} {\rm trellis} \right)}$ is nonseparable due to the convolutional code constraint. Equation (\ref{pdf_coded}) means that we can treat the extrinsic information as
\begin{equation}
{\bf r}_{1x} = {\bf x} + {\bf w}_x,
\end{equation}
where ${\bf w}_x$ is the complex Gaussian noise with the noise level $v_{1x}$. The LLRs of extrinsic information $r$ (any element of ${\bf r}_{1x}$) are given as\footnote{In this study, we consider the convolutional code and 4-QAM modulation with Gray mapping, while the extensions of powerful codes, such as Turbo and LDPC codes with high-order modulations, are straightforward.}
\begin{subequations}\label{SDM}
\begin{align}
{\hat d}_2^{\rm r} & = {\ln}{\left( {\frac{p{\left( \left. d^{\rm r} = 1 \right| r \right)}}{p{\left( \left. d^{\rm r} = 0 \right| r \right)}}} \right)} = \frac{2{\sqrt 2}{\rm Re}{\left( r \right)}}{v_{1x}}, \\
{\hat d}_2^{\rm i} & = {\ln}{\left( {\frac{p{\left( \left. d^{\rm i} = 1 \right| r \right)}}{p{\left( \left. d^{\rm i} = 0 \right| r \right)}}} \right)} = \frac{2{\sqrt 2}{\rm Im}{\left( r \right)}}{v_{1x}}.
\end{align}
\end{subequations}
In (\ref{SDM}), binary bits $d^{\rm r}$ and $d^{\rm i}$ are mapped to the real and imaginary part of symbol $x$, respectively. In Algorithm~1, we use ${\rm SDM}{\left( {\cdot} \right)}$, ${\rm DIL}{\left( {\cdot} \right)}$, and ${\rm IL}{\left( {\cdot} \right)}$ to denote the functions of soft demodulator (\ref{SDM}), deinterleaver, and interleaver, respectively. Moreover, ${\rm APP}{\left( {\cdot} \right)}$ represents the functional of BCJR algorithm, which outputs the LLRs of information bitstream ${\bf b}$ and the encoded bit stream ${\bf s}$. Hence, the posterior pdf of the real part of $x$ can be written as
\begin{subequations}
\begin{align}
p{\left( {\rm Re}{\left( x \right)} = {\frac{1}{\sqrt 2}} \right)} & = {\frac{{\exp}{\left( \frac{{\hat d}_1^{\rm r}}{2} \right)}}{{\exp}{\left( \frac{{\hat d}_1^{\rm r}}{2} \right)} + {\exp}{\left( -\frac{{\hat d}_1^{\rm r}}{2} \right)}}}, \\
p{\left( {\rm Re}{\left( x \right)} = {-}{\frac{1}{\sqrt 2}} \right)} & = {\frac{{\exp}{\left( -\frac{{\hat d}_1^{\rm r}}{2} \right)}}{{\exp}{\left( \frac{{\hat d}_1^{\rm r}}{2} \right)} + {\exp}{\left( -\frac{{\hat d}_1^{\rm r}}{2} \right)}}},
\end{align}
\end{subequations}
whereas that of the imaginary part of $x$ can be obtained similarly with ${\hat d}_1^{\rm i}$. ${\hat d}_1^{\rm r}$ and ${\hat d}_1^{\rm i}$ are the LLRs of binary bits $d^{\rm r}$ and $d^{\rm i}$, respectively. Therefore, the posterior mean and variance of $x$ can be obtained as
\begin{align}
{\hat x}_1 & = {\frac{1}{\sqrt 2}}{\rm tanh}{\left( \frac{{\hat d}_1^{\rm r}}{2} \right)} + {\frac{i}{\sqrt 2}}{\rm tanh}{\left( \frac{{\hat d}_1^{\rm i}}{2} \right)}, \\
v_{1x}^{\rm post} & = 1 - {\frac{{{\rm tanh}^2{\left( \frac{{\hat d}_1^{\rm r}}{2} \right)}} + {{\rm tanh}^2{\left( \frac{{\hat d}_1^{\rm i}}{2} \right)}}}{2}},
\end{align}
where ${\rm tanh}{\left( {\cdot} \right)}$ represents the hyperbolic tangent function. With the above computation details of each estimator, we can perform Algorithm 1 to detect the transmit bitstream of UE.

\begin{remark}
The considered problem (\ref{GLM}) is an instance of generalized linear model (GLM). However, obtaining the MMSE or maximum likelihood solution for GLM is NP-hard. Thus, many efforts are paid to investigate the low-complexity solutions for GLM in the last few years. Research \cite{GEC} proposes the generalized expectation consistent (GEC) algorithm, which is dedicated to the uncoded system, to recover the sparse signal from the general sensing matrix and the quantized output based on the EC inference \cite{EC}. Here, we refer the GEC to the GEC without coding (GEC-U). GEC-U has the same structure as GEC-C, except the computations of the posterior expectation estimator of ${\bf x}$. More precisely, the posterior expectation estimator of ${\bf z}$ and the LMMSE estimator of ${\bf z} = {\bf Ax}$ are the same for GEC-C and GEC-U, and the main difference between the GEC-C and GEC-U is the decoding of ${\bf x}$. In the GEC-U, the posterior pdf of the posterior expectation estimator of ${\bf x}$ is given as
\begin{equation}\label{pdf_uncoded}
p{\left( {{\left. {\bf x} \right|}{{\bf r}_{1x}}{,}{v_{1x}}} \right)} = \frac{{p{\left( {\bf x} \right)}{\exp} {\left( {-}{\frac{{\left\| {{\bf x}{-}{\bf r}_{1x}} \right\|}^2}{v_{1x}}} \right)}}}{{\int}{p{\left( {\bf x} \right)}{\exp}{\left( {-}\frac{{\left\| {{\bf x}{-}{\bf r}_{1x}} \right\|}^2}{v_{1x}} \right)}d{\bf x}}}.
\end{equation}
In the uncoded system, each symbol of transmit signal vector ${\bf x}$ is randomly uniformly generated from a set of constellations, which means that prior distribution $p{\left( {\bf x} \right)}$ is separable. Accordingly, posterior mean ${\bf \hat x}_1$ and variance ${\bf v}_{1x}^{\rm post}$ are computed elementwisely. Posterior pdf (\ref{pdf_coded}) can be seen as a more general case of (\ref{pdf_uncoded}), where the joint prior distribution of ${\bf x}$ is used to compute the local MMSE estimate. We use the convolutional code to construct a desired joint prior distribution, resulting in improved performance than that of the uncoded one. In addition, the computational complexity of the posterior mean and variance is relatively low for the joint prior distribution constructed by the convolutional code.
\end{remark}

\begin{remark}
GEC-C and GEC-U use the low-complexity LMMSE estimator to decouple the transmit symbol of different antennas of UE. GEC-U achieves the Bayesian MMSE performance under the uncorrelated and correlated Rayleigh channels when the system size is large. However, if the channel is ill-conditioned (which means that the condition number of the channel is large), the performance of LMMSE estimator is limited. Moreover, the convergence of GEC-U is poor under the ill-conditioned channels, and the small system size will further aggravate such problem. Hence, we use the coding scheme to mitigate the degradation. In fact, LMMSE estimator works well with the aid of coding scheme even when the channel is ill-conditioned. Considering that the proposed MIMO detector is based on the Bayesian inference, we conjecture that the GEC-C has nearly Bayesian MMSE performance in the large system limit. As a result, the proposed GEC-C has a good performance while keeps the low-complexity. Also, the channel coding provides huge gains in terms of the BER, which results in a substantial power savings of UE.
\end{remark}

\begin{remark}
In the iterative decoding for the coded system, related works often use the belief propagation-based algorithms rather than the expectation propagation-based algorithms. Recently, Ma {\em et al.} proposed a MIMO detection algorithm, which constructs a new estimator for the codeword ${\bf x}$ with the aid of extrinsic message, for the linear coded system based on the expectation propagation \cite{EMA-OAMP}. The proposed EMA-OAMP algorithm has better performance than its belief propagation counterpart, Wang-Poor algorithm \cite{WP}. Although EMA-OAMP algorithm is written in another form, EMA-OAMP algorithm actually follows a regular operations, such as GEC-C and GEC-U: (1) compute the local MMSE estimate with respect to the posterior pdf and joint prior distribution of ${\bf x}$; (2) perform the turbo-type interference cancellation (\ref{ext_m}) and (\ref{ext_v}) for the posterior mean and variance of ${\bf x}$.
\end{remark}

\subsection{Performance Analysis}

\begin{figure}
\centering
\includegraphics[scale = 1]{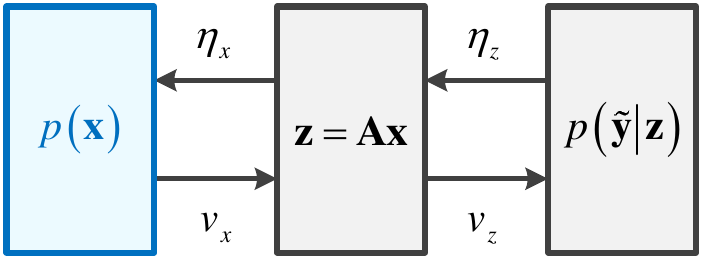}
\caption{Block diagram of the SE equations.}
\label{se_block}
\end{figure}

In the above subsection, we propose the GEC-C for MIMO detection. Although the performance of the GEC-C can be obtained by simulations, understanding the system performance due to the different RIS effects is inconvenient. In this subsection, we present an analysis framework to specify the performance of GEC-C. Under the large system limit when $K$ and $N$ tend to infinity with a fixed ratio ${\frac{K}{N}} = {\alpha}$ and the infinite code length, the behavior of three estimators of GEC-C can be characterized by the transfer functions. In particular, we have the following proposition.

\begin{proposition}\label{SE_coded}
Given the initial conditions ${v_x^0} = 1$ and ${v_z^0} = {\rm tr}{\left( {{\bf A}{\bf A}^H} \right)}/N = P_z$, the saddle point of the free energy of the coded system can be obtained via the following equations
\begin{subequations}
\begin{align}
{\rm 1})~&{\eta_z^{t {+} 1}} = {\left( {\frac{1}{{\zeta}{\left( {{v_z^t}, P_z, v_w, {\Gamma}} \right)}} - {v_z^t}} \right)^{ {-} 1}}, \label{SE_GEC_C_1}\\
{\rm 2})~&{\eta}_x^{t {+} 1} = {\frac{1}{{{v_x^t}}}}{\left( {\frac{1}{{\psi _r}{\left( {{v_x^t}, {\eta_z^{t {+} 1}}, {\bf{A}}} \right)}} - 1} \right)}, \label{SE_GEC_C_2}\\
{\rm 3})~&{v_x^{t {+} 1}} = {{\left( {\frac{1}{{{\rm{mmse}_c}{\left( {\eta _x^{t {+} 1}} \right)}}} - {\eta}_x^{t {+} 1}} \right)}^{ {-} 1}}, \label{v_eta_coded}\\
{\rm 4})~&{v_z^{t {+} 1}} = {\frac{1}{\eta_z^{t {+} 1}}}{\left( {\frac{1}{{\psi _f}{\left( {{v_x^{t {+} 1}}, {\eta_z^{t {+} 1}}, {\bf{A}}} \right)}} - 1} \right)}, \label{SE_GEC_C_4}
\end{align}
\end{subequations}
where the auxiliary functions ${\psi _r}{\left( {{v_x}{,}{\eta_z}{,}{\bf{A}}} \right)}$, ${\psi _f}{\left( {{v_x}{,}{\eta_z}{,}{\bf{A}}} \right)}$, ${\zeta}{\left( {{v_z}{,}P_z{,}v_w{,}{\Gamma}} \right)}$, and ${\rm{mmse}_c}{\left( {\eta_x} \right)}$ are defined as (\ref{SE_channel_reverse}), (\ref{SE_channel_forward}), (\ref{SE_quan}), and (\ref{transfer_coded_app}), respectively. In addition, ${\eta_b^t} = {\delta}{\left( {\eta_x^t} \right)}$, which can be obtained from (\ref{ber_match}), represents the noise precision of real Gaussian corrupted observation ${\bf r}_b = {\bf \bar b} + {\bf w}_t$ with mapping $b_k = 0 \to {\bar b}_k = {-1}$ and $b_k = 1 \to {\bar b}_k = 1$ at the $t$-th iteration. Hence, the BER at $t$-th iteration is given as ${\Phi}{( {-}{\sqrt {{\delta}{\left( {\eta_x^t} \right)}}} )}$.
\end{proposition}
\begin{IEEEproof}
See Appendix~A for the details of the derivation of Proposition~1.
\end{IEEEproof}

Fig.~\ref{se_block} illustrates the block diagram of the above SE equations, which characterize the behavior of three estimators in Fig.~\ref{block_diagram_1_2}. In Proposition~1, the transfer function (\ref{v_eta_coded}) defines the input-output relation of the left block of Fig.~\ref{se_block}. The transfer function (\ref{v_eta_coded}) specifies the behavior of posterior expectation estimator of ${\bf x}$ in the large system limit. Moreover, the transfer function (\ref{SE_GEC_C_1}) corresponds to the posterior expectation estimator of ${\bf z}$, while the transfer functions (\ref{SE_GEC_C_2}) and (\ref{SE_GEC_C_4}) correspond to the LMMSE estimator on the reverse and forward sides, respectively. The iterations of the GEC-C algorithm can be characterized by the iterations of SE equations (transfer functions).

\begin{figure*}
\centering
\resizebox{6.5in}{!}
{
\begin{tabular}{ccc}
\includegraphics*{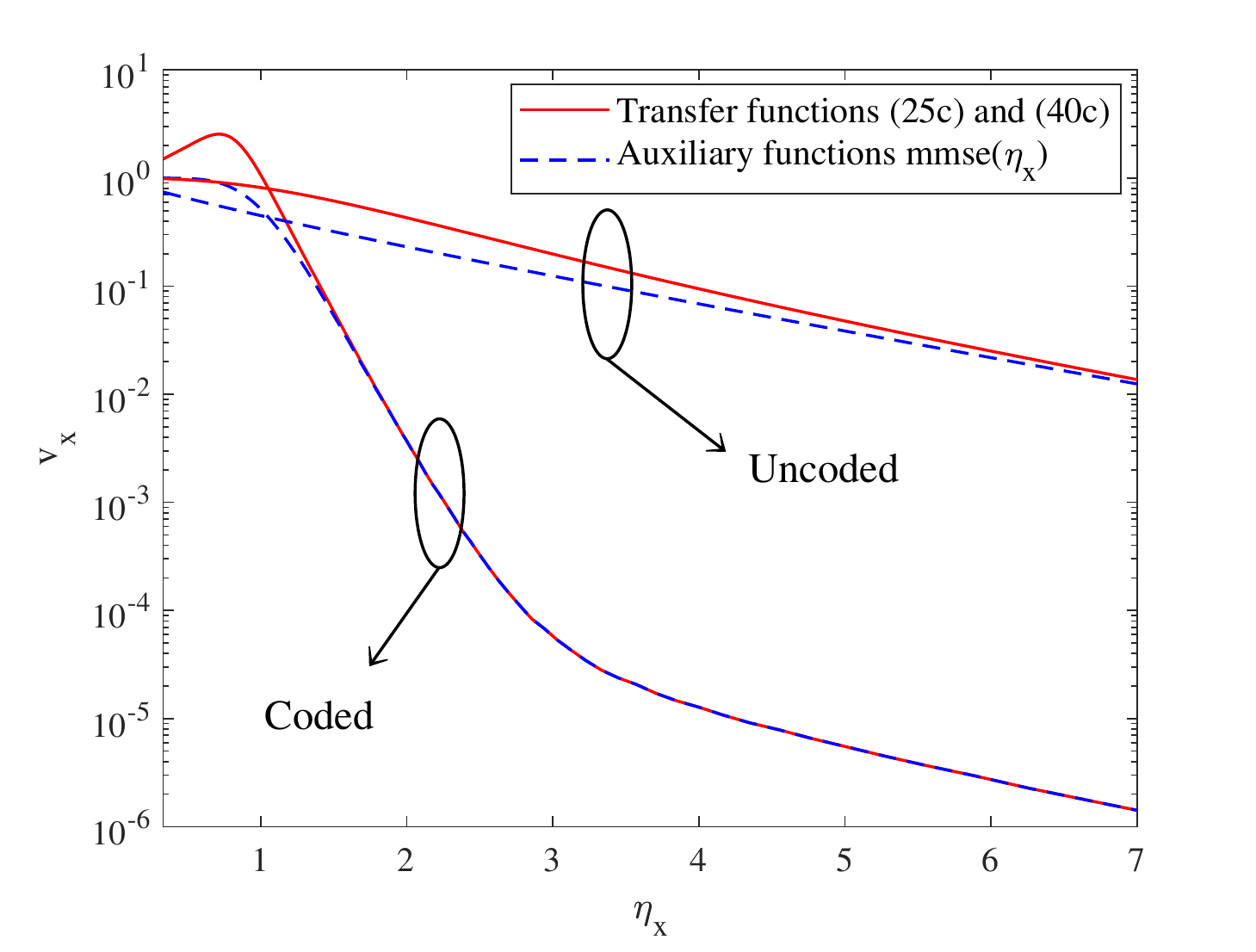} & \includegraphics*{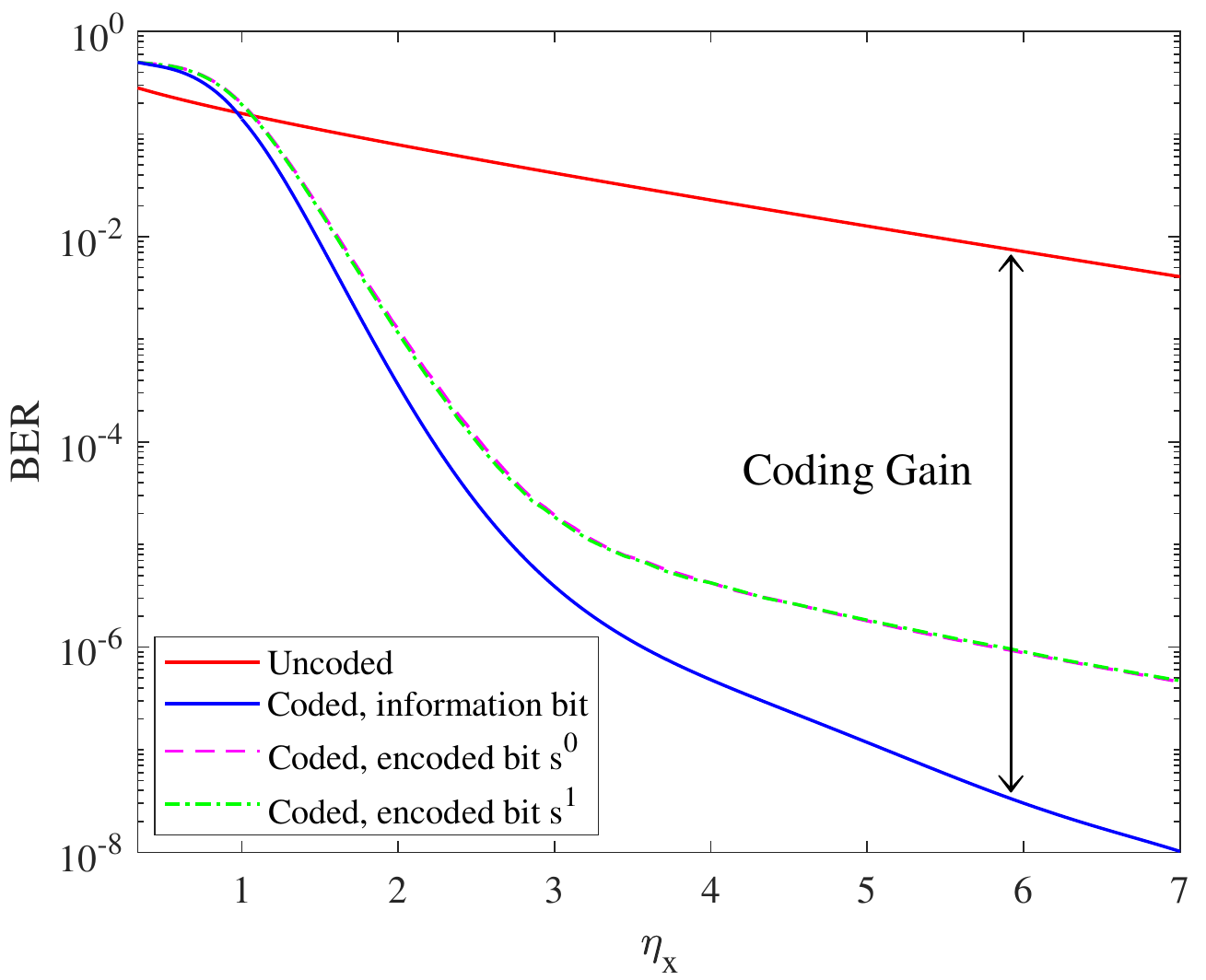} \\
\Large (a) & \Large (b) \\
\end{tabular}
}
\caption{(a) $\eta_x$ versus $v_x$ of transfer functions and auxiliary functions for coded and uncoded systems with rate ${\frac{1}{2}}$ convolutional code and generators $\left( 133, 171 \right)$. (b) $\eta_x$ versus BER for coded and uncoded systems with rate ${\frac{1}{2}}$ convolutional code and generators $\left( 133, 171 \right)$, where encoded bits $s^0$ and $s^1$ are the first and second output bits of the convolutional encoder.}
\label{SE_transfer}
\end{figure*}

\begin{remark}
In fact, both GEC-C and GEC-U are derived based on the EC inference. In Appendix A, we show that the GEC-C and GEC-U has similar analytical framework (i.e., SE equations), except for the transfer function correspond to the posterior expectation estimator of ${\bf x}$. Fig.~\ref{SE_transfer}(a) compares the transfer functions (\ref{v_eta_coded}) and (\ref{v_eta_uncoded}) and the auxiliary functions ${{\rm{mmse}_c}{\left( {\eta_x} \right)}}$ and ${{\rm{mmse}_u}{\left( {\eta_x} \right)}}$. Clearly, ${{\rm{mmse}_u}{\left( {\eta_x} \right)}}$ is a monotonic decreasing function, but ${{\rm{mmse}_c}{\left( {\eta_x} \right)}}$ saturates when $\eta_x$ is small. Consequently, the transfer function of uncoded system (\ref{v_eta_uncoded}) is also a monotonic decreasing function but the transfer function of coded system (\ref{v_eta_coded}) does not. In Section IV, we use this property to explain the BER behavior of GEC-C under the synthetic channel. Moreover, Fig.~\ref{SE_transfer}(b) illustrates the analytical BER performance of coded and uncoded systems under different $\eta_x$. We find that the coding scheme provides huge gains in terms of BER. Besides, the coding system has worse performance than the uncoded one when $\eta_x$ is small. Hence, an appropriate system design should not involve such $\eta_x$ region in the iteration process of SE equations.
\end{remark}

Proposition~1 provides an analytical framework to determine the BER performance of the GEC-C. In the next section, we use the SE equations to evaluate the performance for indoor RIS-assisted mmWave system. As mentioned in Section II, we focus on the {\em (a) RIS effects} and the {\em (b) ADC and phase quantization} of the indoor system.

\section{Performance Analysis for Indoor RIS-assisted MmWave System}

\begin{figure*}
\centering
\includegraphics[scale = 0.8]{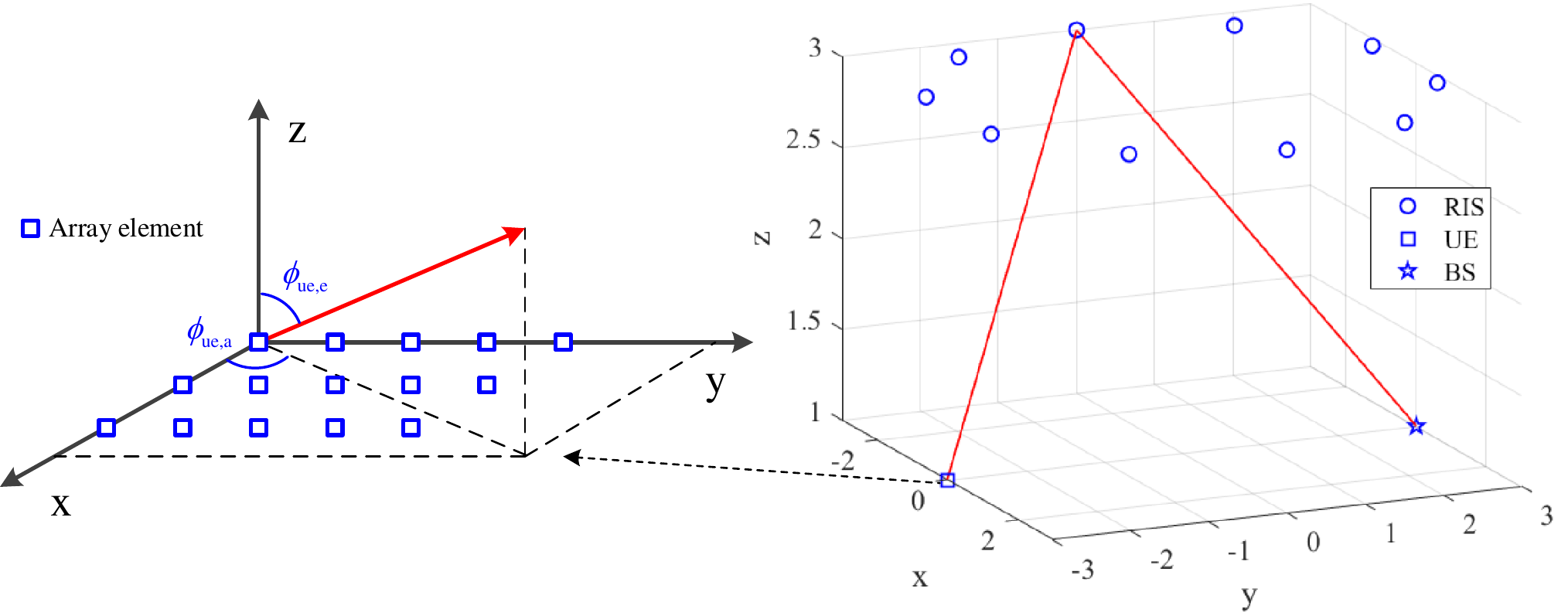}
\caption{3D illustration of indoor RIS-assisted mmWave system.}
\label{indoor}
\end{figure*}

In this section, we consider an indoor RIS-assisted mmWave communication system. The system operates on 28 GHz with a 100 MHz bandwidth, and the power spectral density of the AWGN at the receiver side is $-174$ dBm/Hz. As shown in Fig.~\ref{indoor}, UE and BS are located in a room with a size of $6~{\rm m}{\times}6~{\rm m}{\times}3~{\rm m}$, and a group of $L$ RIS arrays is deployed uniformly around a circle with radius of $3~{\rm m}$ on the surface of ceiling. The antenna arrays are placed on the x-y plane, and Fig.~\ref{indoor} shows the array of UE, where ${\phi}_{\rm ue, a}$ and ${\phi}_{\rm ue, e}$ represent the azimuth and elevation, respectively. For the UE with ULA, if the array is placed along with the x-axis, then the steering vector that corresponds to the $i$th RIS is given as
\begin{align}
& {\bf h}_i^x{\left( {K} \right)} \nonumber\\
= & {\left[ {1{,} e^{j2{\pi}{\frac{d_{\rm ue}}{\lambda}}{\cos}{\phi}_{\rm ue, a}^i{\sin}{\phi}_{\rm ue, e}^i}{,} {\cdots}{,} e^{j2{\pi}{\frac{d_{\rm ue}}{\lambda}}(K{-}1){\cos}{\phi}_{\rm ue, a}^i{\sin}{\phi}_{\rm ue, e}^i}} \right]^H},
\end{align}
while that of the y-axis case is given as
\begin{align}
& {\bf h}_i^y{\left( {K} \right)} \nonumber\\
= & {\left[ {1{,} e^{j2{\pi}{\frac{d_{\rm ue}}{\lambda}}{\sin}{\phi}_{\rm ue, a}^i{\sin}{\phi}_{\rm ue, e}^i}{,} {\cdots}{,} e^{j2{\pi}{\frac{d_{\rm ue}}{\lambda}}(K{-}1){\sin}{\phi}_{\rm ue, a}^i{\sin}{\phi}_{\rm ue, e}^i}} \right]^H}.
\end{align}
The steering vector of URA can be written as
\begin{equation}
{\bf h}_i^{xy}{\left( {K_1{,}K_2} \right)} = {\bf h}_i^x{\left( {K_1} \right)} \otimes {\bf h}_i^y{\left( {K_2} \right)},
\end{equation}
where $K_1 \times K_2 = K$ and $\otimes$ denotes the Kronecker product. Moreover, we have the similar definitions for the steering vectors of BS and RIS, and that of BS is given as
\begin{equation}
{\bf g}_i^{xy}{\left( {N_1{,}N_2} \right)} = {\bf g}_i^x{\left( {N_1} \right)} \otimes {\bf g}_i^y{\left( {N_2} \right)},
\end{equation}
where $N_1 \times N_2 = N$. We use ${\phi}_{\rm bs, a}$ and ${\phi}_{\rm bs, e}$ to represent the azimuth and elevation of BS, respectively. Similar EGC spatial processing such as (\ref{EGC_processing}) can be obtained for the RIS with URA, and it also contributes to enhanced received signal at the BS such as the ULA case. However, different antenna configurations of arrays of UE and BS result in a different synthetic channel, which implies a different BER performance. Also, different locations of UE and BS result in a different wireless channel. As mentioned in Section II, we use the SE equations to analyze the RIS effects of the indoor system. Notably, we assume the spacing of adjacent antennas is half wavelength at the UE and BS, and we use the rate $\frac{1}{2}$ convolutional code with generators $\left( 133, 171 \right)$ throughout the following discussions.

We use the free space path loss model proposed in \cite{PL}. The EGC spatial processing scheme considered at the RIS corresponds to the far-field beamforming case in \cite{PL}, and the path loss factor is given as
\begin{align}
& M \beta_i \nonumber\\
= & M \sqrt{\frac{G_t G_r G d_x d_y \lambda^2 F{\left( \phi_{\rm ue, a}^i, \phi_{\rm ue, e}^i \right)} F{\left( \phi_{\rm bs, a}^i, \phi_{\rm bs, e}^i \right)} A^2}{64 \pi^3 d_{\rm ue, i}^2 d_{\rm bs, i}^2}}. \label{Path_Loss}
\end{align}
In (\ref{Path_Loss}), $F{\left( \phi_{\rm a}, \phi_{\rm e} \right)}$ is the normalized power radiation pattern of the unit cell (reflector) of RIS. Different RIS designs result in different radiation patterns. In this section, we consider the normalized radiation pattern as $F{\left( \phi_{\rm a}, \phi_{\rm e} \right)} = {\rm cos}{\left( \phi_{\rm e} \right)}$. Given the normalized radiation pattern, the antenna gain of the unit cell of RIS can be calculated as
\begin{equation}
G = \frac{4 \pi}{\int_{0}^{2 \pi} \int_{0}^{\frac{\pi}{2}} { {\rm cos} {\left( \phi_{\rm e} \right)} {\rm sin} {\left( \phi_{\rm e} \right)} } d{\phi_{\rm e}} d{\phi_{\rm a}} } = 4.
\end{equation}
Furthermore, the antenna gains of UE and BS are $G_t = G_r = 1$. The size of each unit cell along the x-axis is $d_x$, and that along the y-axis is $d_y$, which are usually of subwavelength scale ranging from $0.1\lambda$ to $0.5\lambda$, where $\lambda$ is the wavelength. Moreover, $A$ is the reflection coefficient. In this section, we have $d_x = d_y = 0.1 \lambda$, and $A = 0.9$. We use the RIS array (URA) with 90 rows and columns; hence, $M = 8100$. Finally, $d_{\rm ue, i}$ denotes the distance between the UE and the $i$-th RIS, whereas $d_{\rm bs, i}$ represents the distance between the $i$-th RIS and the BS.

\subsection{Overview of the SE Process}

\begin{figure}
\centering
\includegraphics[scale = 0.6]{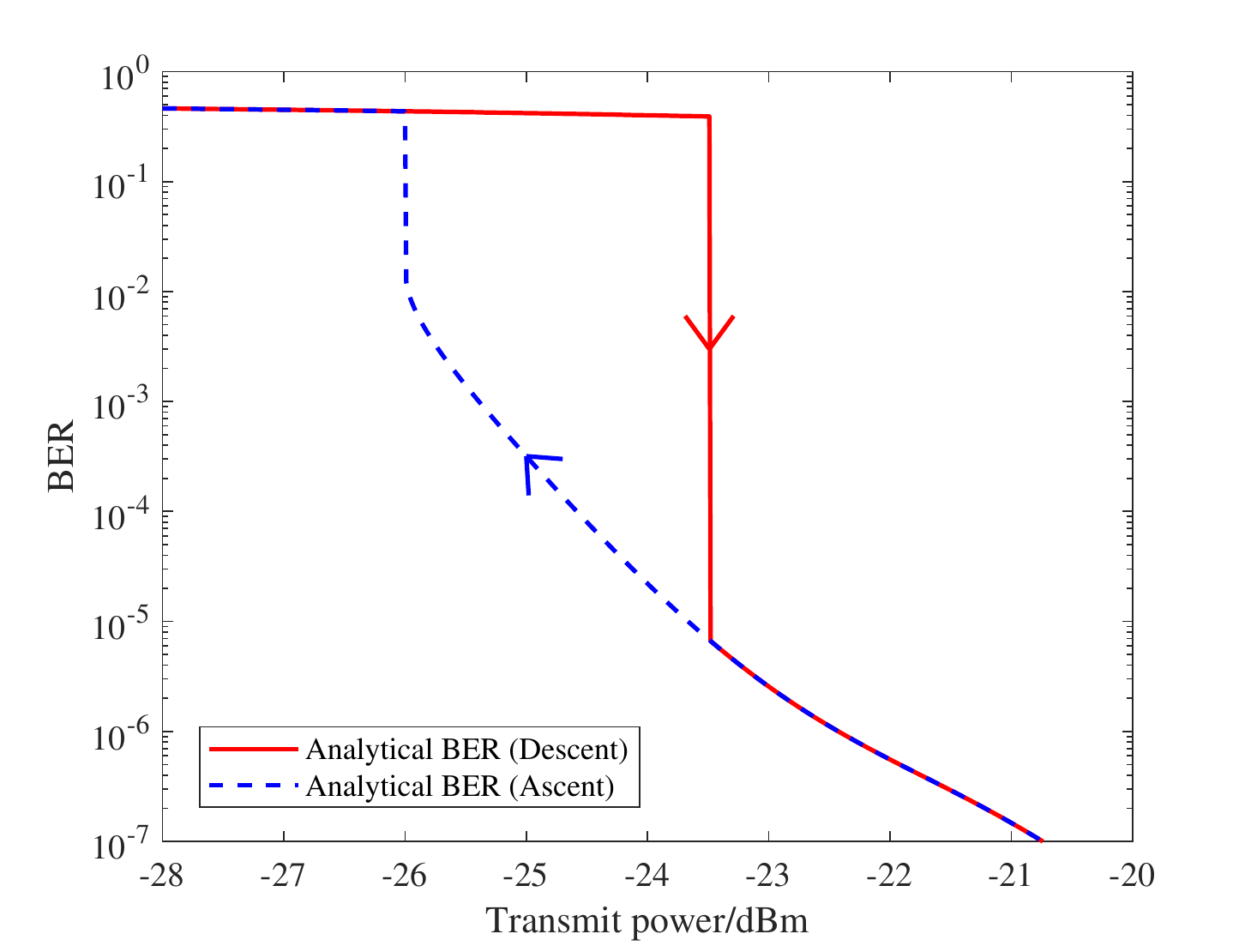}
\caption{Analytical BER performances obtained from SE equations with descent and ascent processes.}
\label{ascent}
\end{figure}

Fig.~\ref{ascent} shows the BER performances of a RIS-assisted system with ULA ($K_1 = 8$ and $N_1 = 16$), where $L = 10$ and ${\rm B} = 3$. The solid line in Fig.~\ref{ascent} is the BER obtained from Proposition~2. We find that the BER drops suddenly when the signal-to-noise ratio (SNR) is larger than the threshold. In fact, the iterations of the SE equations can be viewed as a descent process along with the curve of transfer functions (\ref{v_eta_coded}) and (\ref{v_eta_uncoded}) for coded and uncoded systems, respectively. According to Remark~4, the transfer function of coded system (\ref{v_eta_coded}) is not a monotonic decreasing function, so it requires an ``initial impetus'' to push the evolution state ${\left( {\eta_x}, {v_x} \right)}$ crosses the ``peak''. Once the ``initial impetus'' exceeds the threshold, the evolution state will drop along with the slope and stop at a position far from the ``peak''. Therefore, the BER of a coded system behaves like a diode. By contrast, the uncoded system does not have such phenomenon since the transfer function of uncoded system (\ref{v_eta_uncoded}) is a monotonic decreasing function. Such phenomenon is also related to relevant studies in statistical physics.

The SE equations claim that the BER of the coded system is high when the SNR is smaller than the threshold. However, SE analysis is derived in the large system limit and will deviate from the simulations when the system size is small. The BER of a small system will remain low even when the SNR is smaller than the threshold. We refer the Proposition~2 as the SE equations with descent process, and we introduce the SE equations with ascent process to fill this gap. In the SE equations with ascent process, we provide a large value to ${\eta}_x^1$ to force the evolution state cross the ``peak’’. Then, the evolution state will rise along with the curve of transfer function (\ref{v_eta_coded}). In Fig.~\ref{ascent}, dashed line represents the BER obtained from the SE equations with ascent process. The BER of the ascent process decreases linearly at the medium SNR. Besides, solid and dashed lines coincide with each other at the other SNR regions. SE equations with descent and ascent processes provide a general framework to evaluate the performance of GEC-C. In statistical physics, this ``trick'' is used to evaluate the state of the spin glasses \cite{Replica}.

\subsection{RIS Effects}

\begin{figure*}
\centering
\resizebox{6.5in}{!}
{
\begin{tabular}{ccc}
\includegraphics*{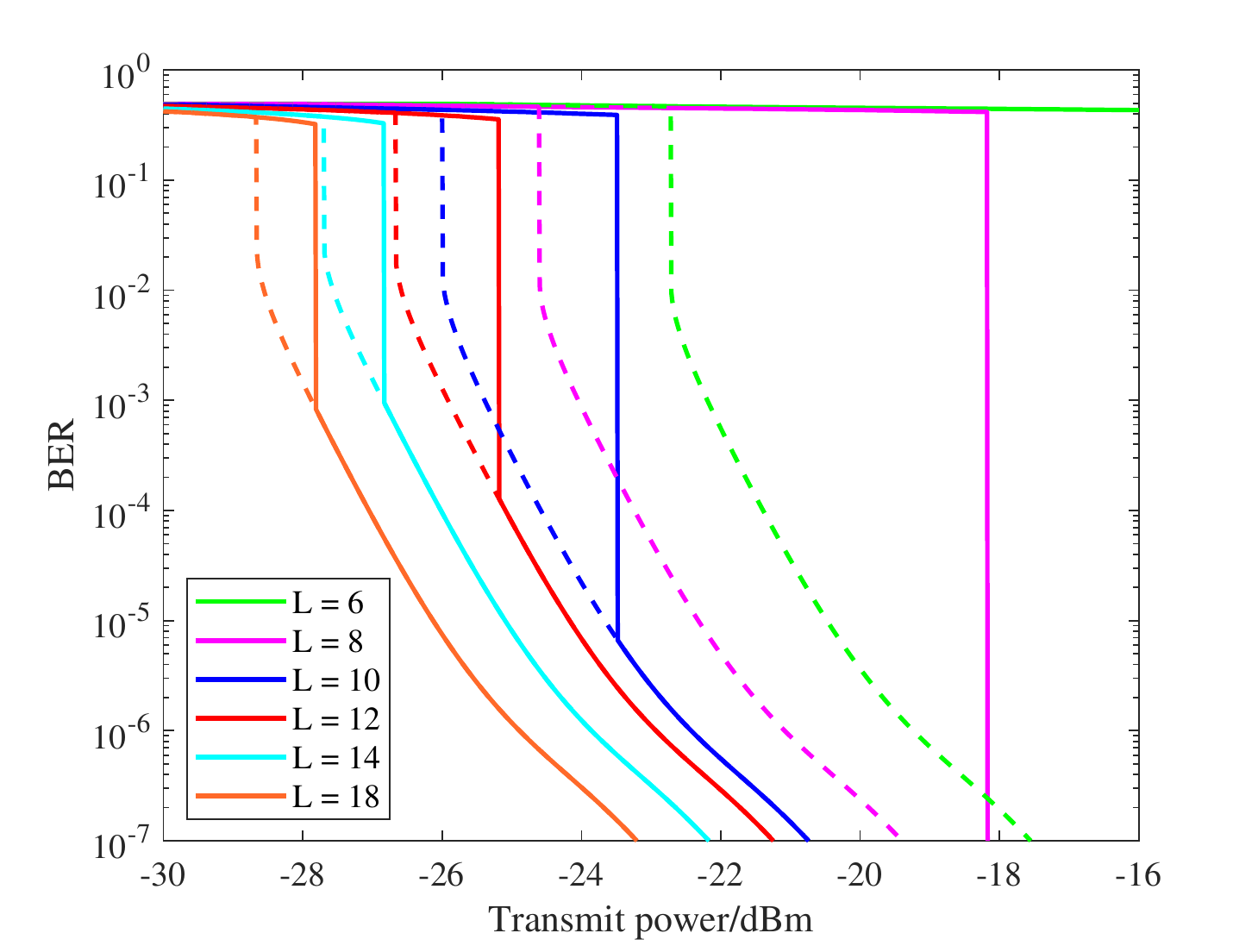} & \includegraphics*{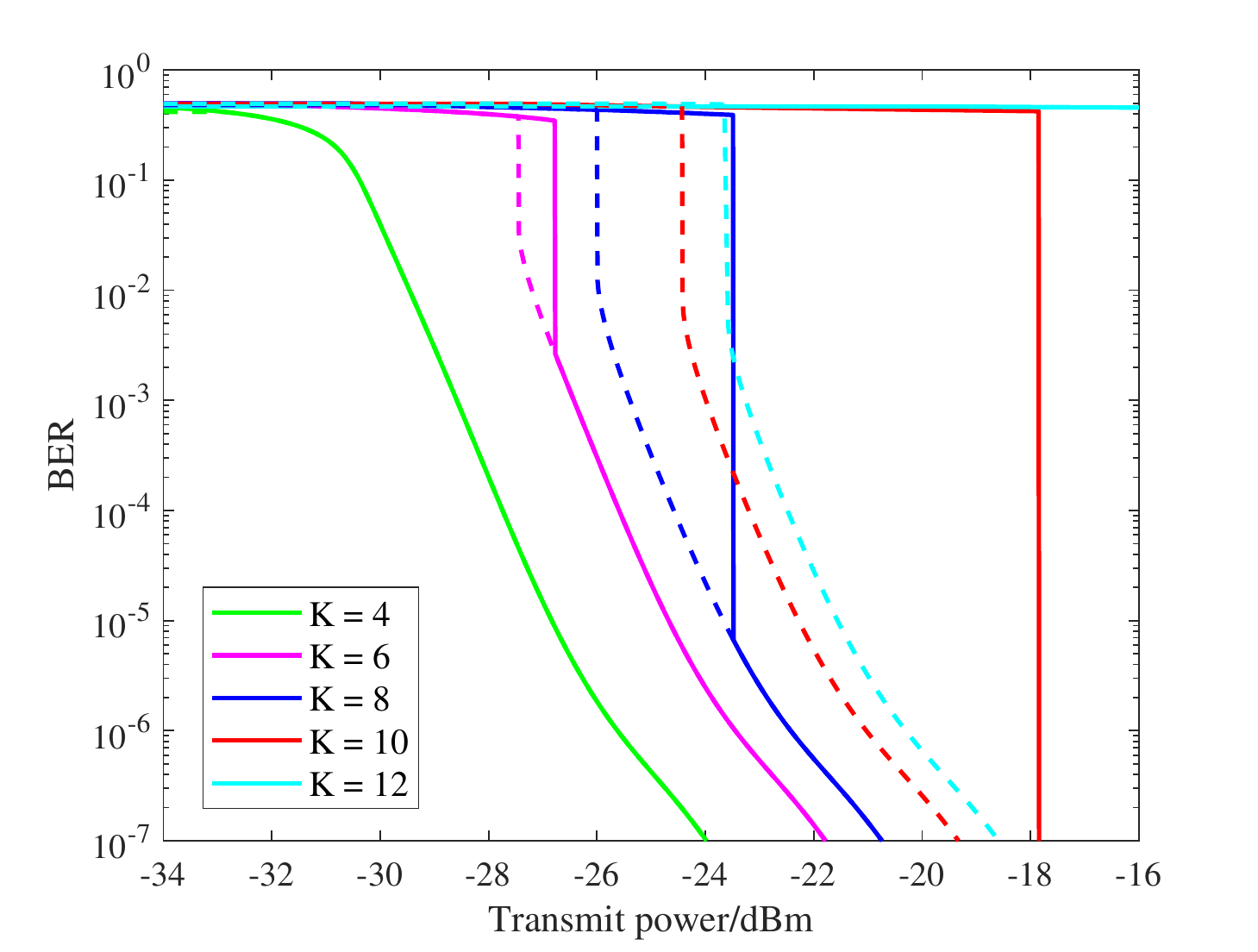} \\
\Large (a) & \Large (b) \\
\end{tabular}
}
\caption{BER performances under the different (a) number of RIS arrays and (b) number of UE antennas.}
\label{SE_L_K}
\end{figure*}

\begin{figure*}
\centering
\resizebox{6.5in}{!}
{
\begin{tabular}{ccc}
\includegraphics*{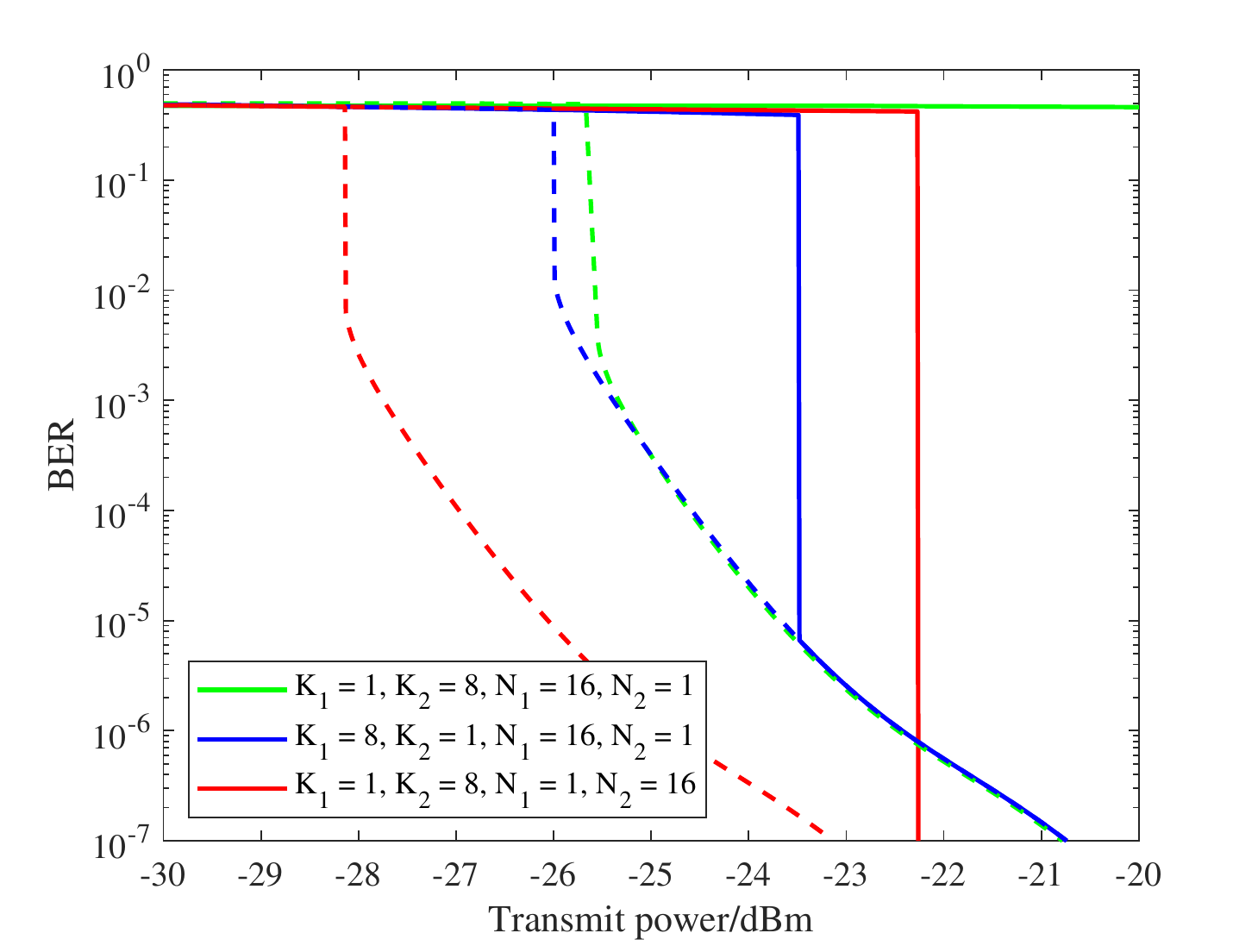} & \includegraphics*{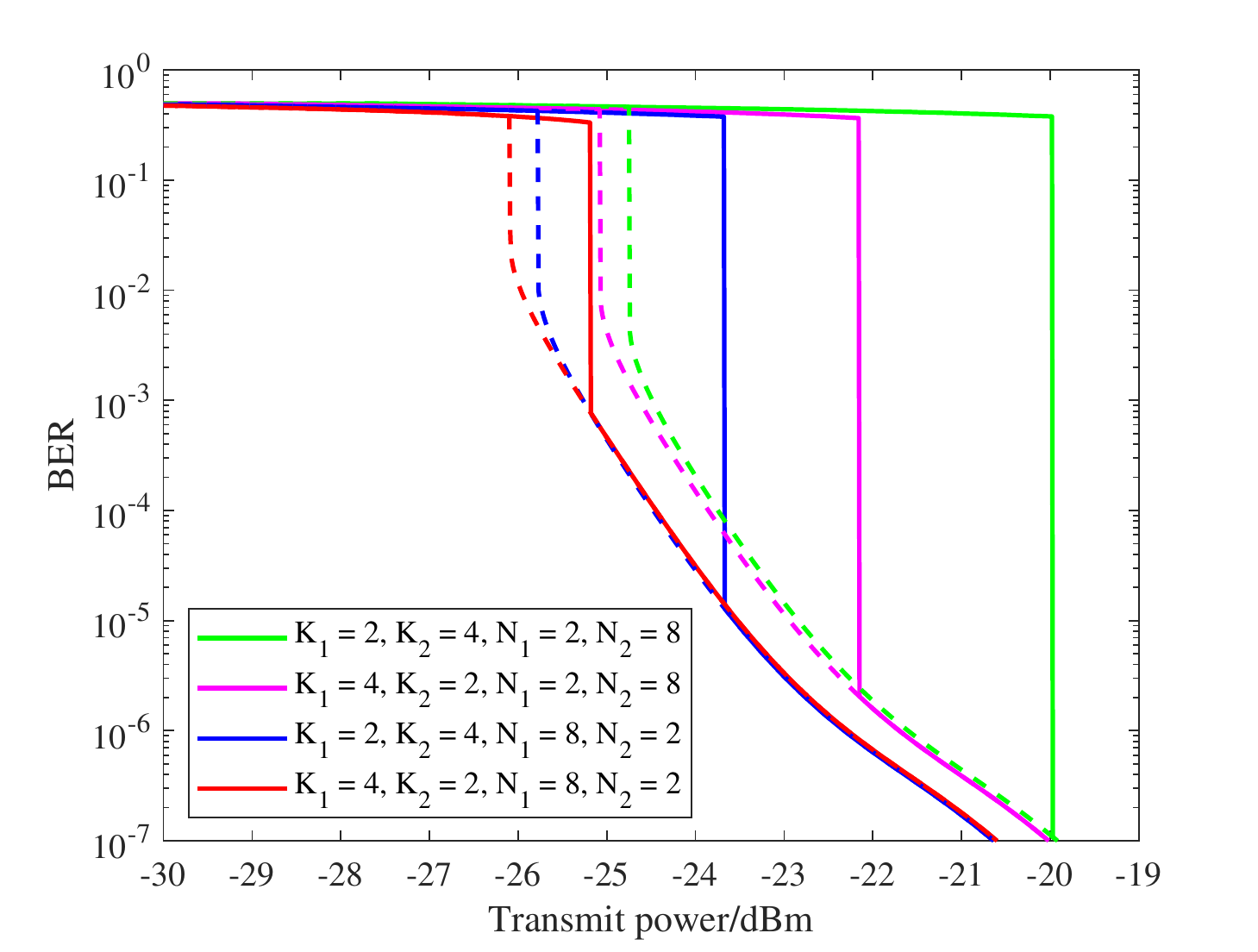} \\
\Large (a) & \Large (b) \\
\end{tabular}
}
\caption{BER performances under the different (a) ULA configurations and (b) URA configurations.}
\label{SE_ULA_URA}
\end{figure*}

In this subsection, we discuss the RIS effects of indoor system. Consider a RIS-assisted system with ULA ($K_1 = 8$ and $N_1 = 16$) and ${\rm B} = 3$. Based on the SE equations, Fig.~\ref{SE_L_K}(a) shows the BER performances under the different numbers of RIS arrays. The BER drops immediately when SNR is larger than the threshold, especially for a small number of RIS arrays. A small number of RIS arrays means poor spatial diversity and large condition number. According to (\ref{SE_channel_reverse}), an unevenly distributed set of singular values causes smaller ``initial impetus'' $\eta_x^1$ compared with the evenly distributed one. Hence, channel matrix with a less number of RIS arrays requires a large SNR to escape the trap. The threshold depends on the characteristics of the channel. By increasing the number of RIS arrays, we can improve the performance of the medium BER region. Fig.~\ref{SE_L_K}(b) shows the BER performances under the different numbers of UE antennas, where $L = 10$ and ${\rm B} = 3$. The condition number of the channel matrix decreases when the ratio $\frac{K}{N}$ decreases, and the trap effect nearly vanishes for $K = 4$. By contrast, the system with less spatial diversity only works well in the low BER region. We find that the supportable number of UE antennas ($K$) depends on the number of strong singular values. For example, the channel matrix with $K = 14$ has six strong singular values only. The equivalent number of independent data streams is $K/2 = 7$ because we use the rate $\frac{1}{2}$ convolutional code. Under such conditions, the system cannot work well because the number of independent data streams exceeds the number of strong singular values. Given $L = 10$, the maximum supportable number of UE antennas is $K = 12$, which equates to a spectral efficiency of 12 bps/Hz.

In the above discussion, we consider the case where the UE and the BS are ULA. However, different antenna configuration has a different BER performance because the synthetic channel is fully characterized by the steering vectors. Fig.~\ref{SE_ULA_URA}(a) illustrates the BER performances under the different ULA configurations, where $L = 10$ and ${\rm B} = 3$. The case of $K_1 = 8$, $K_2 = 1$, $N_1 = 1$, and $N_2 = 16$ is omitted because it has nearly the same performance as the case of $K_1 = 1$, $K_2 = 8$, $N_1 = 16$, and $N_2 = 1$. We find that the different layout of ULA affects the performance significantly, where the case of $K_1 = 1$, $K_2 = 8$, $N_1 = 1$, and $N_2 = 16$ has nearly 2.5 dB gain compared with other cases. Moreover, making two ULAs parallel is better than orthogonal for exploiting spatial diversity. Fig.~\ref{SE_ULA_URA}(b) shows the BER performances under the different URA configurations. Compared with ULA, URA is relatively insensitive in the low BER region because it exploits the diversities of both azimuth and elevation. However, the performances of different URA cases are inferior to the ULA case of $K_1 = 1$, $K_2 = 8$, $N_1 = 1$, and $N_2 = 16$. The possible values of elevation are limited in the $\left[ 0, \pi/2 \right)$ region while the possible values of azimuth lie in the $\left[ 0, \pi \right)$ region because the RIS arrays are deployed on the surface of the ceiling in the considered indoor scene. Therefore, the diversity of elevation is limited relative to the diversity of azimuth, resulting in an inferior performance for URA.

\begin{figure*}
\centering
\resizebox{6.5in}{!}
{
\begin{tabular}{ccc}
\includegraphics*{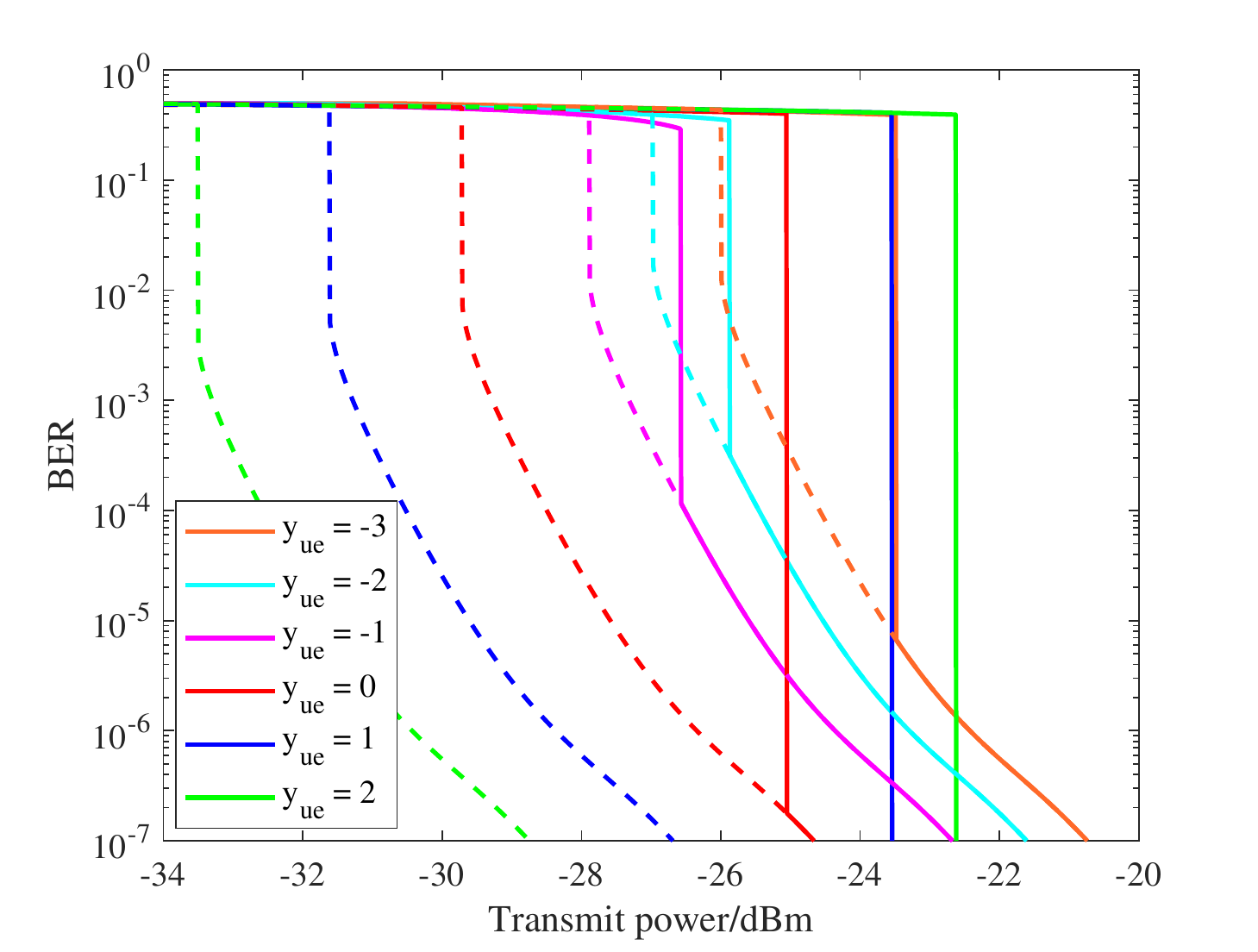} & \includegraphics*{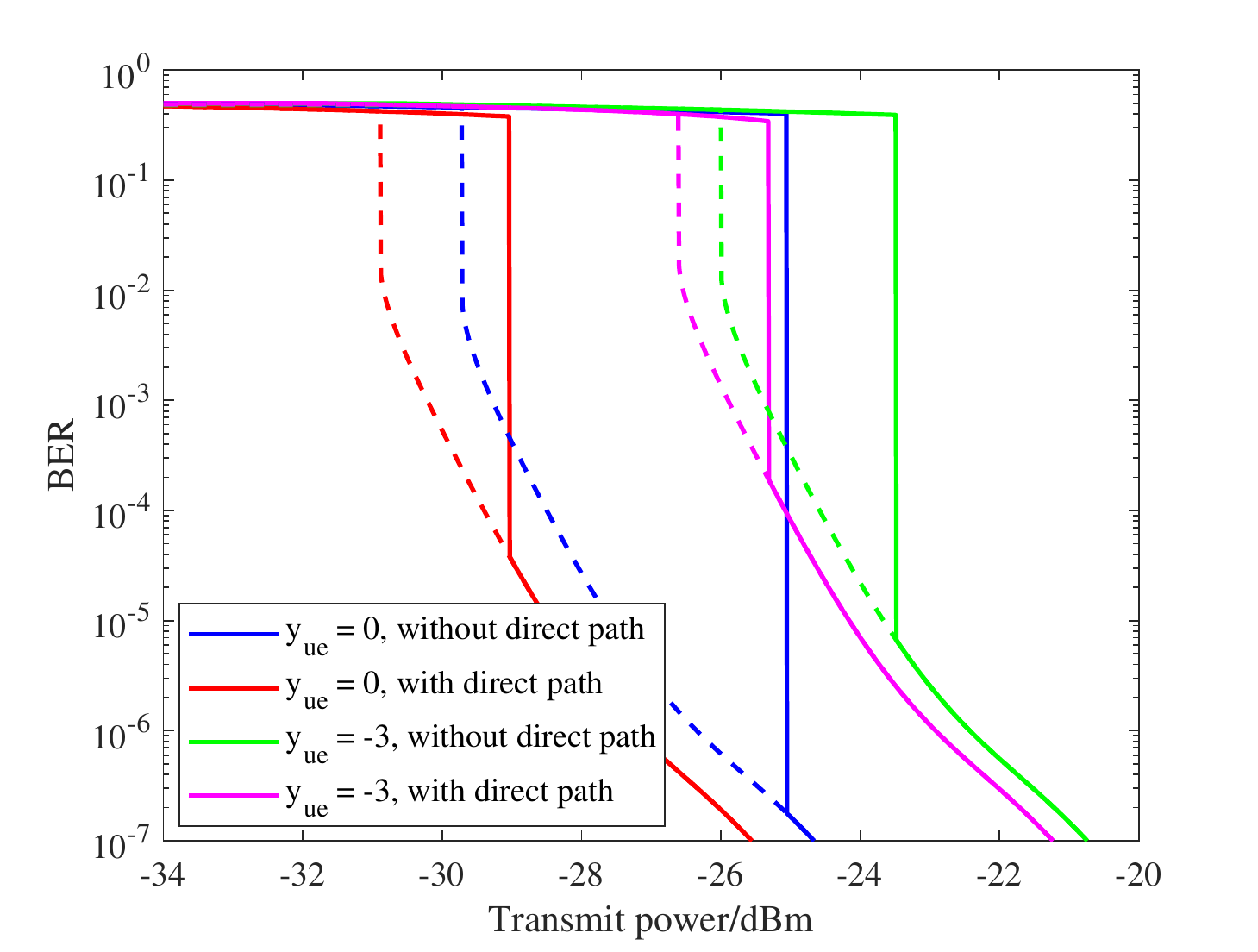} \\
\Large (a) & \Large (b) \\
\end{tabular}
}
\caption{BER performances (a) with different locations of UE and (b) with and without direct path.}
\label{SE_D}
\end{figure*}

\begin{figure*}
\centering
\resizebox{6.5in}{!}
{
\begin{tabular}{ccc}
\includegraphics*{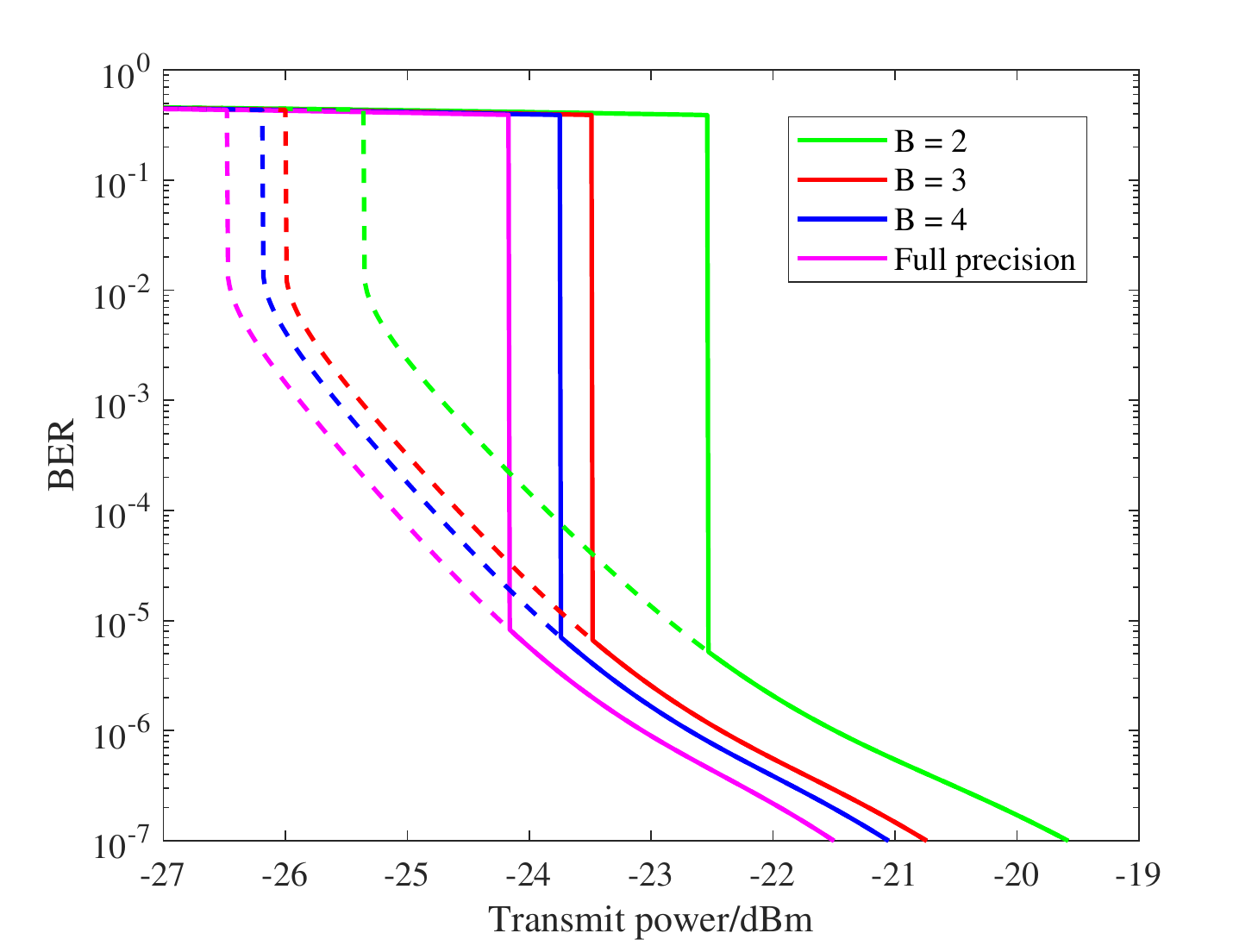} & \includegraphics*{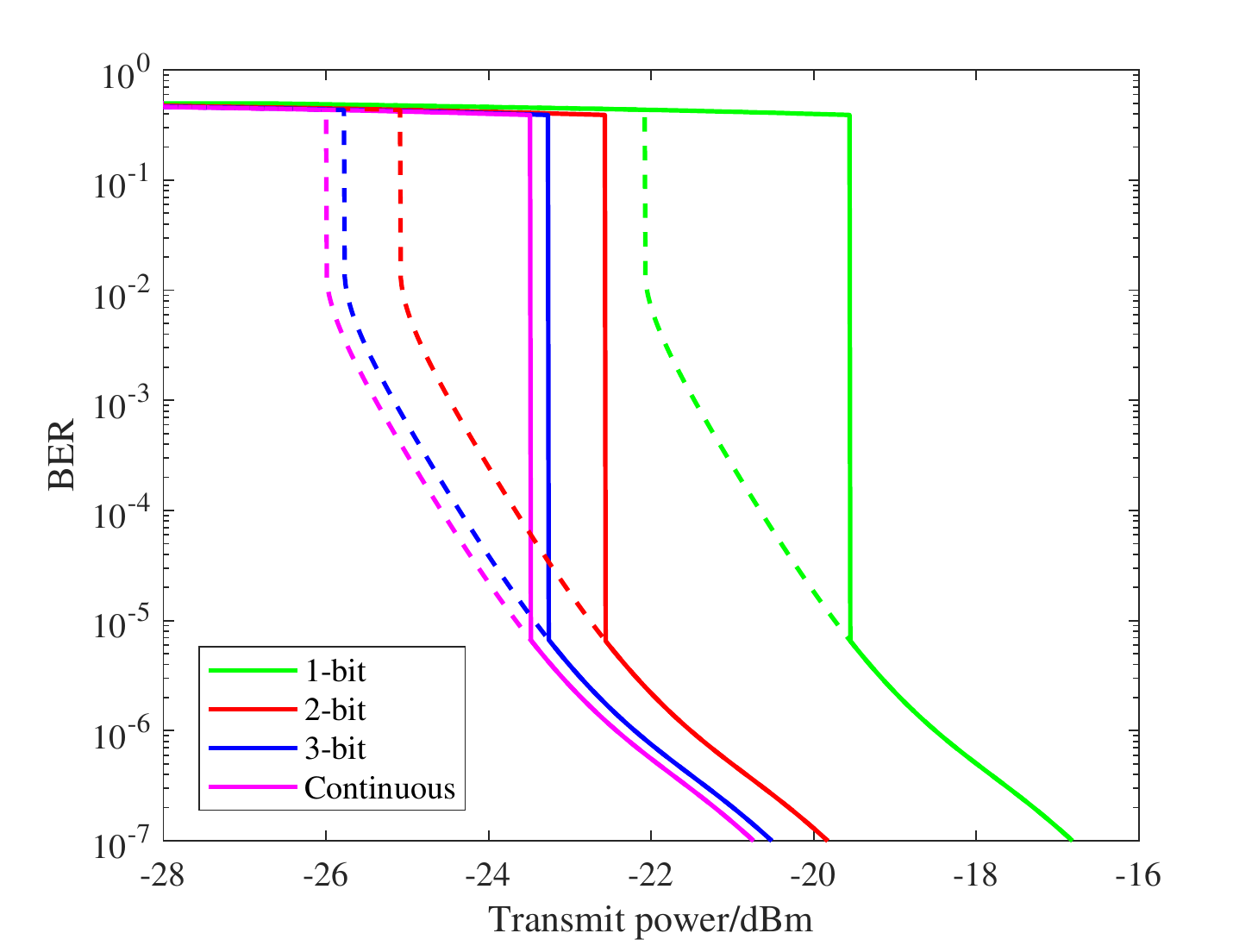} \\
\Large (a) & \Large (b) \\
\end{tabular}
}
\caption{BER performances under the different (a) number of quantization bits and (b) discrete phase shifts.}
\label{SE_B_P}
\end{figure*}

The location of UE also has an important influence. Fig.~\ref{SE_D}(a) illustrates the BER performances with different locations of UE (different y coordinates of UE) under the ULA ($K_1 = 8$ and $N_1 = 16$), where $L = 10$ and ${\rm B} = 3$. We find that the system performance increases with decreasing distance between UE and BS. However, the performance suffers degradation in the medium BER region when UE moves toward BS. When the distance between the UE and BS is small, the signal reflected from the far side RIS arrays suffers larger attenuation compared with that of the near side ones. The spatial diversity is less under such condition. In general, the deployment of RIS arrays such as Fig.~\ref{indoor} also has benefits. This deployment offers a large spatial diversity under the various conditions because the RIS arrays are spatially separated. In the above analysis, we assume that the direct path disappeared due to the obstacles. In Fig.~\ref{SE_D}(b), we consider the mmWave system with a direct path, and we use the free space path loss model as
\begin{equation}
\beta_{\rm los} = \sqrt{\frac{G_t G_r \lambda^2}{16 \pi^2 d_{\rm ue, bs}^2}},
\end{equation}
where $d_{\rm ue, bs}$ is the distance between the UE and the BS. We find that the direct path provides gains relative to the system without a direct path.

\begin{remark}
The mmWave system should be carefully designed if the direct path exists. If the gain of the direct path is stronger than that of the multipath provided by the RIS, the channel matrix will tend to be a rank 1 matrix, which cannot offer a multiplexing gain. To offer a comparable gain as the direct path, the size and number of reflectors of the RIS should be large enough according to (\ref{Path_Loss}). Nevertheless, the direct path leads to the question of how to design the transmission scheme on the basis of the channel characteristics. In traditional mmWave communication, we prefer the LoS transmission because it provides a large gain. A typical transmission scheme uses beamforming technology to transmit few data streams along the direct path. By contrast, the RIS-assisted mmWave system has several multipaths. Without the direct path, we can obtain the multiplexing gain easily because the gains of different multipaths are comparable. The RIS array without EGC spatial processing still provides a multipath because it is a highly reflective material. However, a multiplexing gain with a strong direct path is relatively difficult to obtain because MIMO detection is challenging under the channel matrix with a large condition number. Hence, the RIS-assisted mmWave system may prefer the case without a direct path if the multiplexing gain is large enough.
\end{remark}

\subsection{ADC and Phase Quantization}

\begin{figure*}
\centering
\resizebox{6.5in}{!}
{
\begin{tabular}{ccc}
\includegraphics*{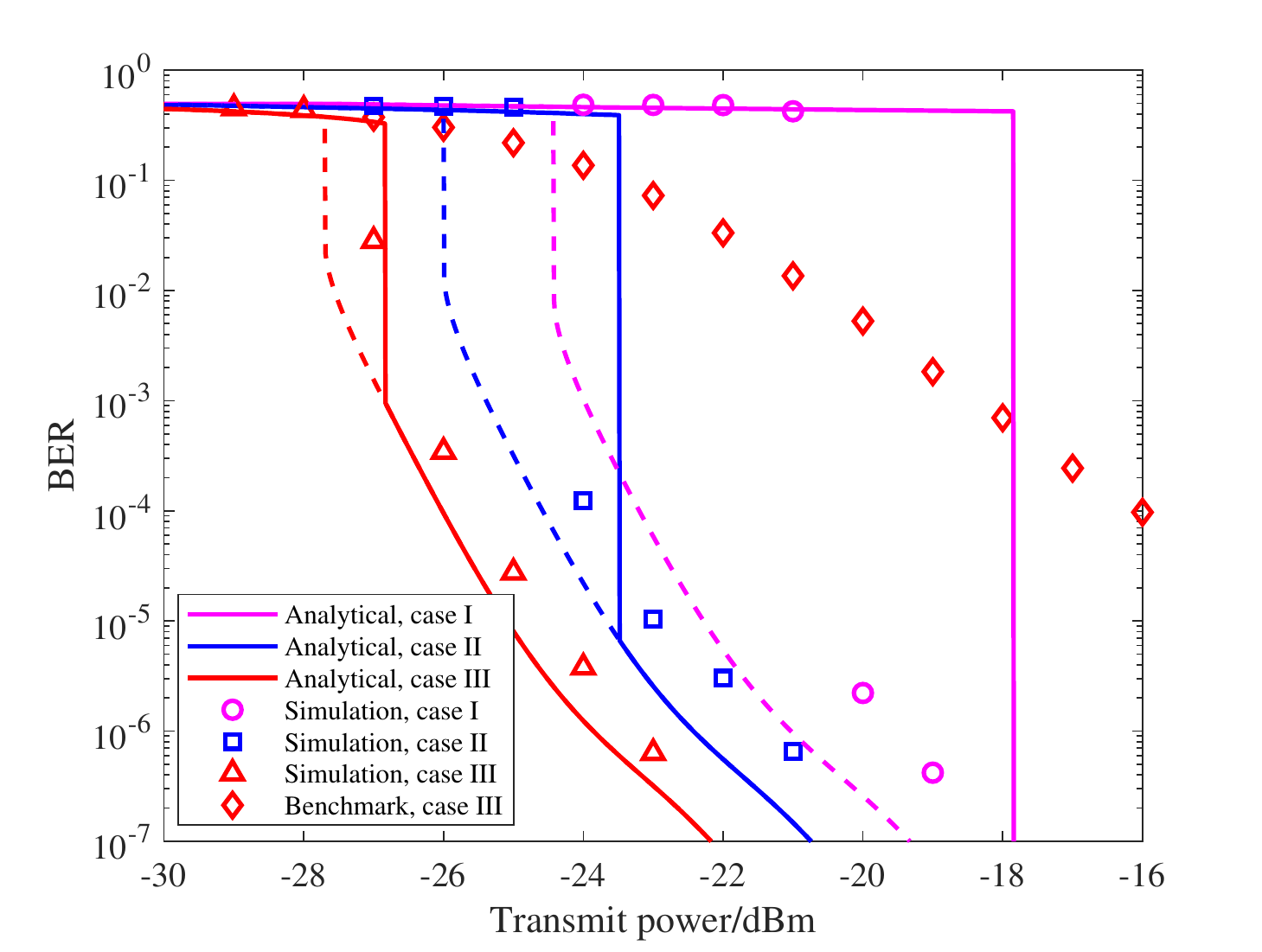} & \includegraphics*{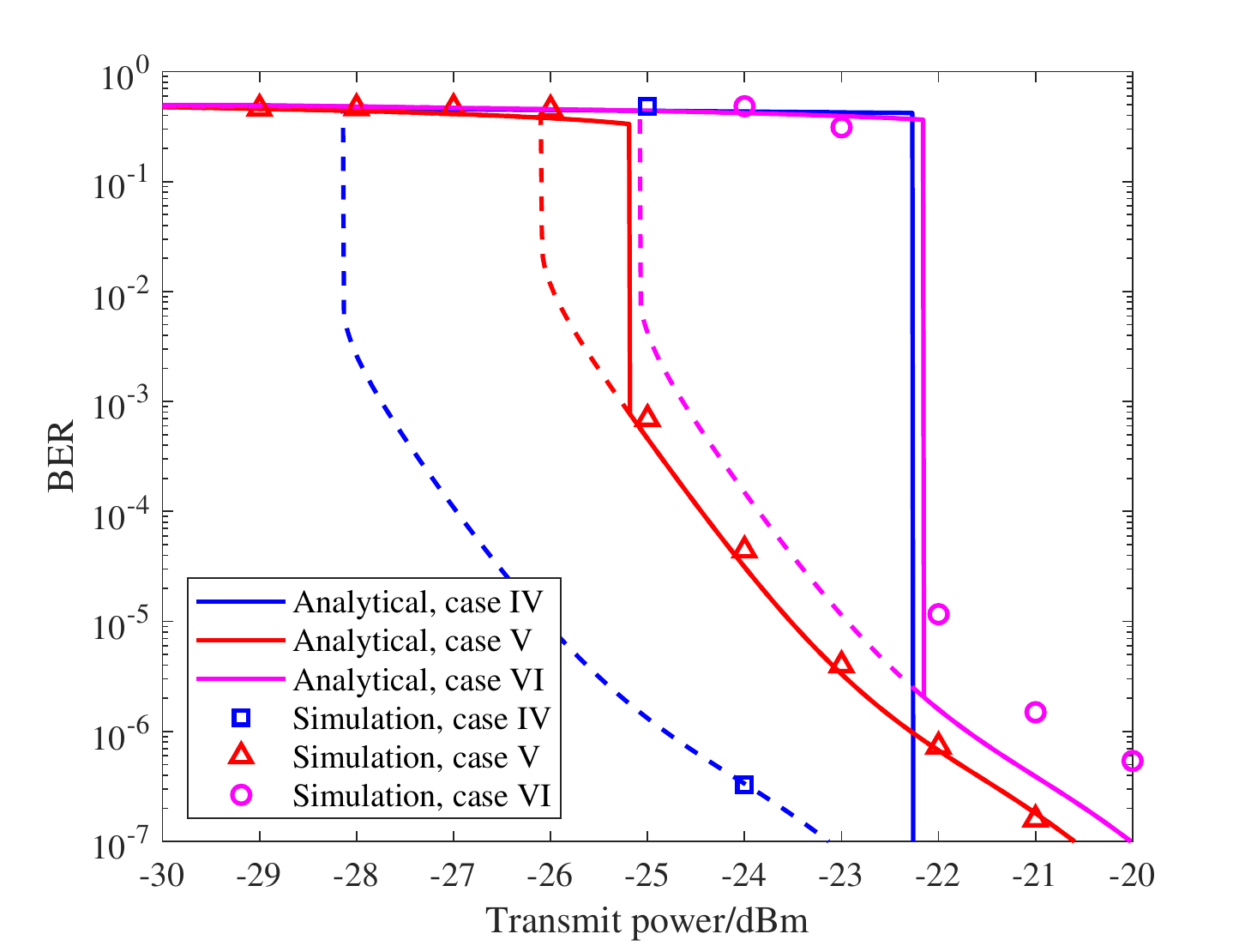} \\
\Large (a) & \Large (b) \\
\end{tabular}
}
\caption{Analytical and simulated BER performances under the different (a) numbers of RIS arrays and UE antennas and (b) different antenna configurations. Case I: $K = 10$, $L = 10$; case II: $K = 8$, $L = 10$; case III: $K = 8$, $L = 14$; case IV: $K_1 = 1$, $K_2 = 8$, $N_1 = 1$, $N_2 = 16$; case V: $K_1 = 4$, $K_2 = 2$, $N_1 = 8$, $N_2 = 2$; case VI: $K_1 = 4$, $K_2 = 2$, $N_1 = 2$, $N_2 = 8$.}
\vspace{-0.2cm}
\label{Simulation_1}
\end{figure*}

ADC is very expensive in the mmWave wireless communication systems because of the high sampling rate. However, we find that the low-precision ADC, that is, 2-bit to 4-bit, does not cause severe performance loss. Fig.~\ref{SE_B_P}(a) shows the BER performances under the different numbers of quantization bits, where the system with ULA ($K_1 = 8$ and $N_1 = 16$) and $L = 10$ is considered. High-precision ADC only provides few gains when ${\rm B} > 3$. Hence, using few-bit ADCs, such as 3-bit ADC, is reasonable in the RIS-assisted mmWave system with GEC-C detector. Previous studies that used expectation propagation-based algorithms also showed that the low-precision ADC is sufficient in the MIMO systems \cite{CE_2,JCD}. Linear detectors, such as maximum ratio combining detector and zero-forcing detector, have limited performance when the low-precision ADC is used.

\begin{figure}
\centering
\includegraphics[scale = 0.6]{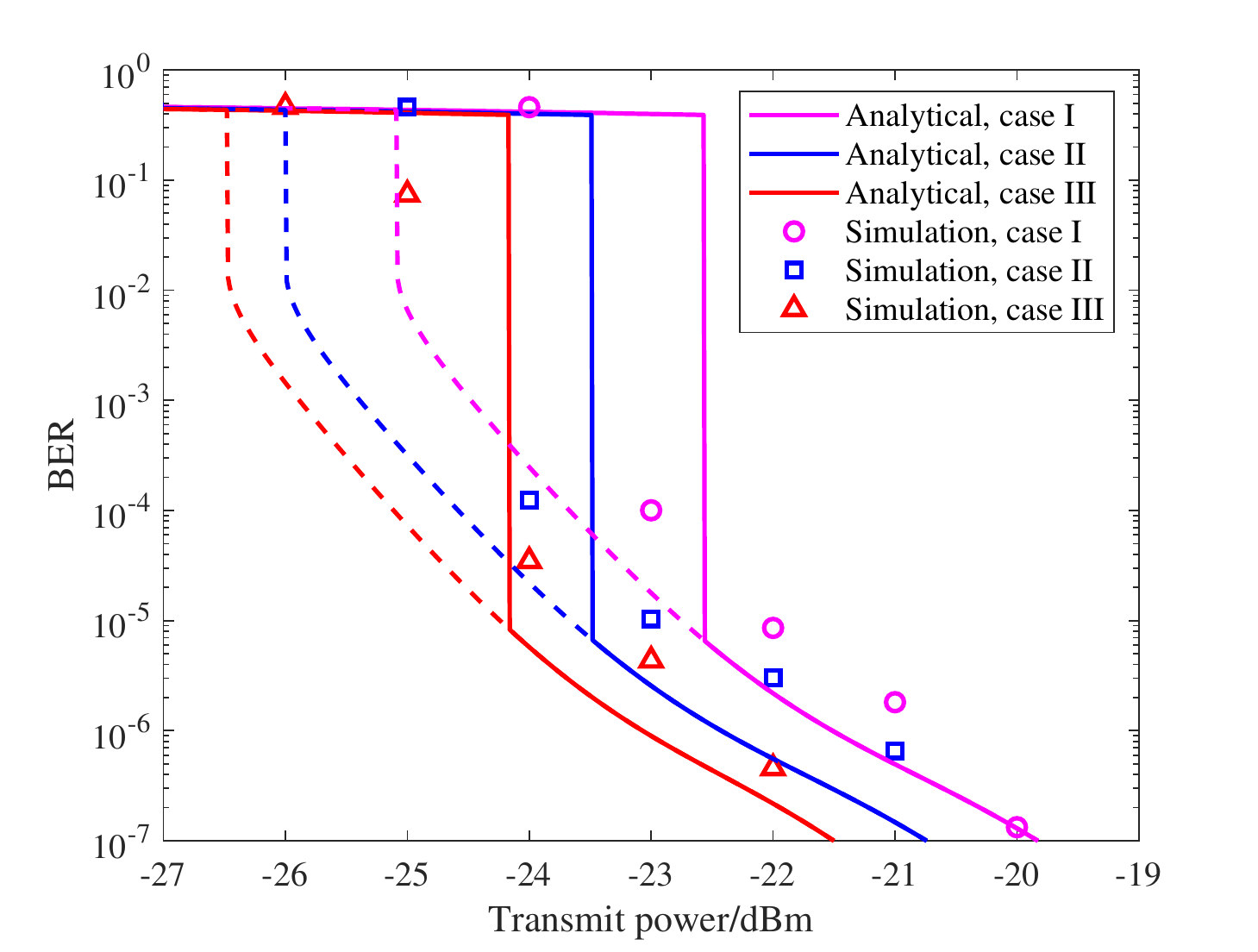}
\caption{Analytical and simulated BER performances under the different hardwares. Case I: 3-bit ADC and 2-bit discrete phase shift; case II: 3-bit ADC and continuous phase shift; case III: full-precision ADC and continuous phase shift.}
\label{Simulation_2}
\end{figure}

In the above analysis, we assume the phase shift is continuous, which enables the perfect EGC spatial processing. However, RIS array with continuous phase shift may cause expensive hardware cost. Therefore, performance losses caused by the discrete phase shift should be investigated. Here, we consider the system with ULA ($K_1 = 8$ and $N_1 = 16$), $L = 10$ and ${\rm B} = 3$. The phase shift value of each reflector is obtained by mapping the continuous phase shift to the nearest points of discrete phase shift, which are uniformly placed. Fig.~\ref{SE_B_P}(b) shows the BER performances under the different discrete phase shifts. We find that the discrete phase shift does not cause severe performance loss. 3-bit discrete phase shift is sufficient to achieve the performance of the continuous phase shift. Moreover, 2-bit discrete phase shift only causes nearly 1 dB loss. These results are also effective for various system settings. Consequently, the proposed RIS-assisted system with GEC-C detector can adopt the low-cost RIS arrays with small number of phase quantization.

\subsection{Simulations versus Analytical Results}

In this part, we use simulations to verify the SE equations with descent and ascent processes. We use the rate $\frac{1}{2}$ convolutional code with generators $\left( 133, 171 \right)$ and the codeword length is 19200. Fig.~\ref{Simulation_1}(a) illustrates the BER performances of the different numbers of RIS arrays and UE antennas with ULA (the arrays of UE and BS are placed along with the x-axis) and ${\rm B} = 3$. The simulations coincide with the analytical results. A large number of RIS array provides not only power gains but also spatial diversity. We find that the SE equations with ascent process are very helpful to evaluate the system performance. The SE equations with descent process show that the system cannot work in the BER of $10^{-7}$ when $K = 10$ and $L = 10$. However, the SE equations with descent process are not very accurate because of the small size of the MIMO system. The system can work well under the SNRs that are relatively smaller than the threshold. We use a traditional detector as the benchmark. According to the additive quantization noise model [50, pp. 125-133], we use the LMMSE estimate to obtain the estimate of $\bf x$. Then, we obtain the encoded bitstream by using hard decision and a deinterleaver. Lastly, the information bitstream can be obtained by using the Viterbi decoder. Fig.~\ref{Simulation_1}(a) shows the performance of the benchmark of case III. The proposed detector outperforms the benchmark detector. Fig.~\ref{Simulation_1}(b) shows the BER performances of different antenna configurations. In fact, case I is better than cases II and III. However, case I only works well in the low BER region. With the help of proposed analytical tool, we can design the appropriate SNR to ensure effective transmission. Fig.~\ref{Simulation_2} shows BER performances under the different hardwares. Clearly, the low-cost hardware does not cause severe performance degradation.

\section{Conclusion}

In this study, we proposed a novel low-cost mmWave system with the aid of several RIS arrays. Through linear spatial processing, these RIS arrays formed a synthetic channel, which enables MIMO transmission. In particular, the synthetic channel had large power gain but possibly less spatial diversity. Hence, we investigated the MIMO detection of a coded system to mitigate the less spatial diversity and reduce the transmit power of UE. A low-complexity MIMO detector, called GEC-C, was developed based on the Bayesian inference. To evaluate the BER performance of the proposed MIMO detector, we presented SE equations with descent and ascent processes as an analytical tool. Through leveraging by the SE equations, we studied the (a) RIS effects and (b) ADC and phase quantization for the indoor system. A large number of RIS arrays provided remarkable gains in terms of BER by enhancing the spatial diversity and received power. Moreover, URA was inferior to ULA due to the limited diversity of elevation. However, since the ULA only exploited the diversity of azimuth, it was relatively sensitive to the orientations of the arrays of UE and BS. The proposed deployment of RIS arrays was robust for different UE locations since the RIS arrays are deployed separately to ensure the spatial diversity. We discussed the effect of direct path and showed that the system should be carefully designed to exploit the multiplexing gain. In addition, our results showed that low-cost hardware, such as the 3-bit ADCs of the BS and the 2-bit uniform discrete phase shift of the RIS arrays, only moderately reduces the performance.

\begin{appendices}

\section{Derivation of Proposition~1}

Considering that GEC-C is related to GEC-U, we conduct the analysis of GEC-U at first. In \cite{GEC}, the authors use replica method, which is derived from statistical physics \cite{Replica}, to analyze the performance of GEC-U. In fact, the performance of GEC-U can be more straightforwardly characterized by the SE equations in the large system limit, when $K$ and $N$ tend to infinity with a fixed ratio ${\frac{K}{N}} = {\alpha}$. We define several auxiliary functions at first. The auxiliary function corresponds to the posterior expectation estimator of ${\bf x}$ is defined as
\begin{equation}\label{transfer_uncoded}
{{\rm{mmse}_u}{\left( {\eta_x} \right)}} = {\int} {{\left( {1 - {\rm tanh}{\left( x \right)}} \right)}^2{\mathcal N}{\left( {x; {\eta_x}, {\eta_x}} \right)}dx}.
\end{equation}
The auxiliary functions correspond to the LMMSE estimator are given as
\begin{align}
{\psi _r}{\left( {{v_x}{,}{\eta_z}{,}{\bf A}} \right)} & = {\rm E}{\left\{ {\frac{1}{{1 + {v_x}{\eta_z}{\lambda}^2}}} \right\}}, \label{SE_channel_reverse}\\
{\psi _f}{\left( {{v_x}{,}{\eta_z}{,}{\bf A}} \right)} & = 1 - {\alpha}{\left( 1 - {\rm E}{\left\{ {\frac{1}{{1 + {v_x}{\eta_z}{\lambda}^2}}} \right\}} \right)}, \label{SE_channel_forward}
\end{align}
where the expectations in (\ref{SE_channel_reverse}) and (\ref{SE_channel_forward}) are taken over the singular values $\lambda$ of channel matrix ${\bf A}$. For example, if ${\bf A}$ has singular values $\lambda_1, \cdots, \lambda_K$, then we have
\begin{equation}
{\rm E}{\left\{ {\frac{1}{{1 + {v_x}{\eta_z}{\lambda}^2}}} \right\}} = \frac{1}{K} \sum\nolimits_{j = 1}^{K} {\frac{1}{{1 + {v_x}{\eta_z}{\lambda}_j^2}}}.
\end{equation}
The auxiliary function corresponds to the posterior expectation estimator of ${\bf z}$ is written as
\begin{equation}\label{SE_quan}
{{\zeta}{\left( {{v_z}, P_z, v_w, {\Gamma}} \right)}} = {\sum\limits_{b = 1}^{{2^{{\rm B}}}}} {\int {\frac{{{{\left( {{\Psi'}{\left( {b; \sqrt {\frac{{{P_z} - {v_z}}}{2}}z, \frac{{v_w + {v_z}}}{2}} \right)}} \right)}^2}}}{{2{\Psi} {\left( {b; \sqrt {\frac{{{P_z} - {v_z}}}{2}}z, \frac{{v_w + {v_z}}}{2}} \right)}}}{\rm D}z}},
\end{equation}
where
\begin{align}
{\Psi}{\left( {b; a, c^2} \right)} & = {\Phi}{\left( {\frac{{r_b} - a}{c}} \right)} - {\Phi}{\left( {\frac{{r_{b{-}1}} - a}{c}} \right)}, \\
{\Psi}'{\left( {b; a, c^2} \right)} & = -\frac{1}{\sqrt {2 \pi c^2}} {\left( e^{ - {\frac{{\left( {r_b} - a \right)}^2}{2c^2}} } - e^{ - {\frac{{\left( {r_{b - 1}} - a \right)}^2}{2c^2}} } \right)},
\end{align}
and ${\rm D}z = {\frac{1}{\sqrt{2{\pi}}}}{\exp}{( -{\frac{z^2}{2}} )} dz$ is the real Gaussian integration measure. In (\ref{SE_quan}), ${\left\{ {r_b}, b = 0, 1, {\cdots}, 2^{\rm B} \right\}} \in {\Gamma}$ denotes the thresholds of ${\rm B}$-bit real-valued quantizer. Based on the above definitions, we have the following proposition for the uncoded system with GEC-U.
\begin{proposition}\label{SE_uncoded}
Given the initial conditions ${v_x^0} = 1$ and ${v_z^0} = {\rm tr}{\left( {{\bf A}{\bf A}^H} \right)}/N = P_z$, the saddle point of the uncoded system can be obtained via the following equations
\begin{subequations}
\begin{align}
{\rm 1})~&{\eta_z^{t {+} 1}} = {\left( {\frac{1}{{\zeta}{\left( {{v_z^t}, P_z, v_w, {\Gamma}} \right)}} - {v_z^t}} \right)^{ {-} 1}}, \label{SE_GEC_U_1}\\
{\rm 2})~&{\eta}_x^{t {+} 1} = {\frac{1}{{{v_x^t}}}}{\left( {\frac{1}{{\psi _r}{\left( {{v_x^t}, {\eta_z^{t {+} 1}}, {\bf{A}}} \right)}} - 1} \right)}, \label{SE_GEC_U_2}\\
{\rm 3})~&{v_x^{t {+} 1}} = {{\left( {\frac{1}{{{\rm{mmse}_u}{\left( {\eta _x^{t {+} 1}} \right)}}} - {\eta}_x^{t {+} 1}} \right)}^{ {-} 1}}, \label{v_eta_uncoded}\\
{\rm 4})~&{v_z^{t {+} 1}} = {\frac{1}{\eta_z^{t {+} 1}}}{\left( {\frac{1}{{\psi _f}{\left( {{v_x^{t {+} 1}}, {\eta_z^{t {+} 1}}, {\bf{A}}} \right)}} - 1} \right)}, \label{SE_GEC_U_4}
\end{align}
\end{subequations}
where $\eta_x^t$ represents the noise precision of complex Gaussian noise corrupted observation ${\bf r}_{1{\bf x}} = {\bf x} + {\bf w}_t$ at the $t$-th iteration. Hence, the BER at $t$-th iteration is given as ${\Phi}{( {-}{\sqrt {\eta_x^t}} )}$.
\end{proposition}
\begin{IEEEproof}
See \cite{GEC} for the details of the derivation of Proposition~2.
\end{IEEEproof}

\begin{figure*}
\centering
\resizebox{6.5in}{!}
{
\begin{tabular}{ccc}
\includegraphics*{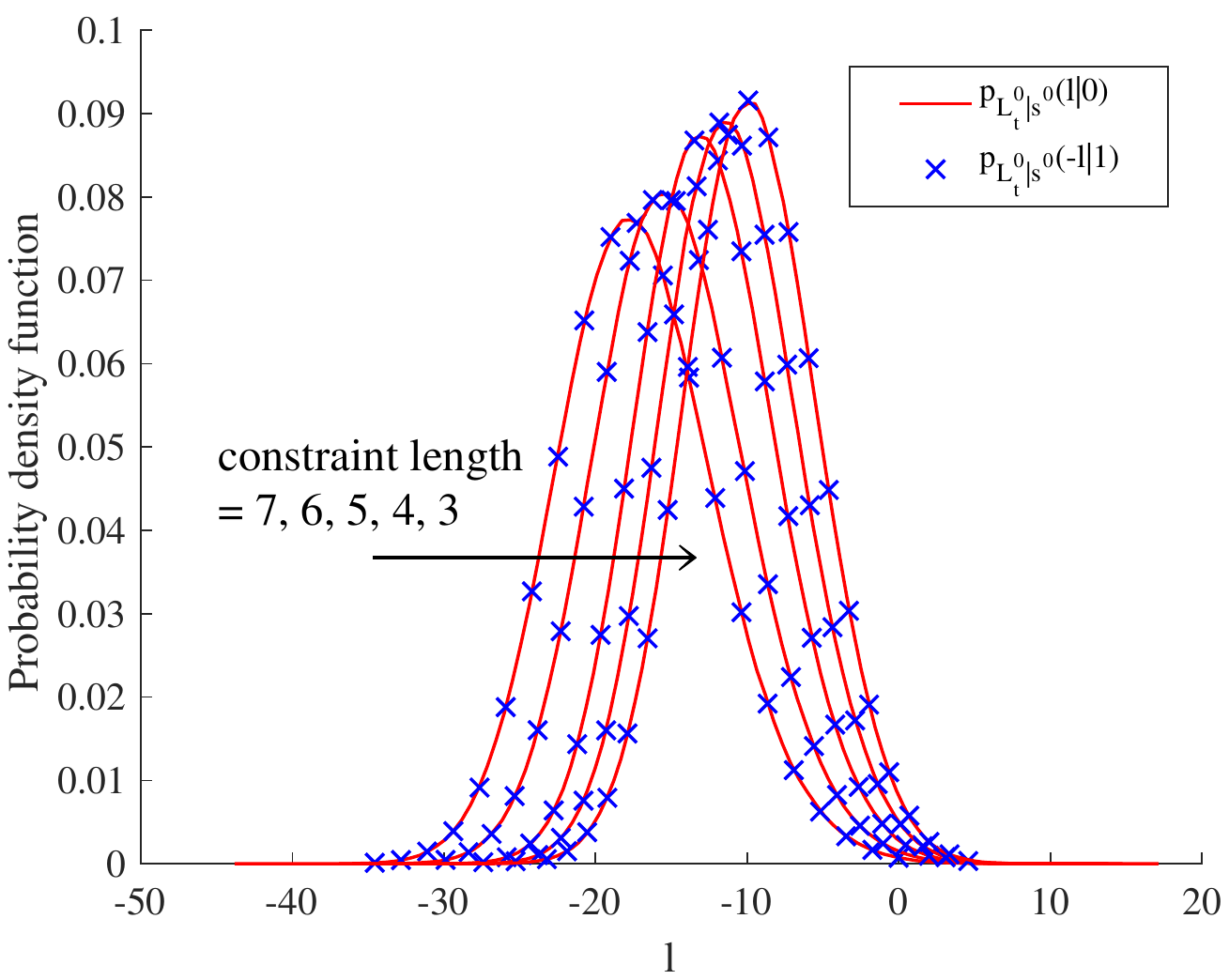} & \includegraphics*{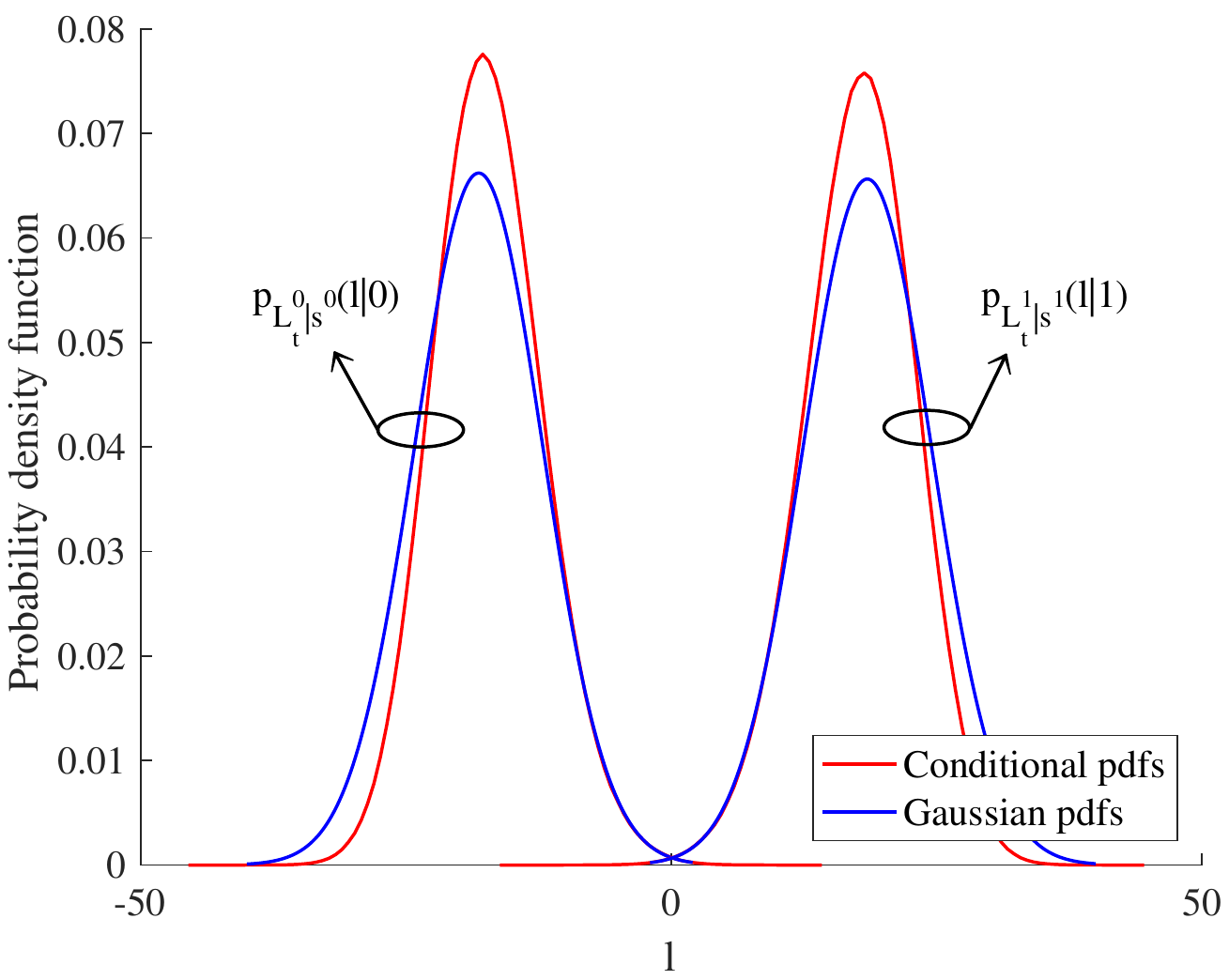} \\
\Large (a) & \Large (b) \\
\end{tabular}
}
\caption{(a) Conditional pdfs of $s^0$ of rate $\frac{1}{2}$ optimal convolutional codes \cite{opt_cc} with ${\eta}_x^t = 2$ and different constraint length. The encoded bit $s^0$ is the first output bit of the convolutional encoder. (b) Conditional pdfs and its Gaussian approximations of rate ${\frac{1}{2}}$ convolutional code with generators $\left( 133, 171 \right)$ and ${\eta}_x^t = 2$.}
\label{pdf_1_2}
\end{figure*}

\begin{table*}
\centering
\caption{Conditional pdfs of information bit and encoded bits.}
\begin{tabular}{|c|c|c|c|c|c|c|}
\hline
  & \multicolumn{2}{c|}{Information bit $b$} & \multicolumn{2}{c|}{Encoded bit $s^0$} & \multicolumn{2}{c|}{Encoded bit $s^1$} \\ \hline
Conditional pdf & $p_{\left. {L_t^b} \right| b}{\left( {l | 0} \right)}$ & $p_{\left. {L_t^b} \right| b}{\left( {l | 1} \right)}$ & $p_{\left. {L_t^0} \right| {s^0}}{\left( {l | 0} \right)}$ & $p_{\left. {L_t^0} \right| {s^0}}{\left( {l | 1} \right)}$ & $p_{\left. {L_t^1} \right| {s^1}}{\left( {l | 0} \right)}$ & $p_{\left. {L_t^1} \right| {s^1}}{\left( {l | 1} \right)}$ \\ \hline
\end{tabular}
\end{table*}

SE equations of GEC-C can be developed by replacing the transfer function of the uncoded prior distribution (\ref{v_eta_uncoded}) with the transfer function of the coded prior distribution (\ref{v_eta_coded}). Assuming a large system limit with infinite code length, we can use the density evolution analysis to obtain ${\rm{mmse}_c}{\left( {\eta_x} \right)}$. For convenience, the proposed analysis framework is dedicated to the rate $\frac{1}{2}$ convolutional code, and the extensions of different code rates are straightforward. For rate $\frac{1}{2}$ convolutional code, an information binary bit $b_k$ is mapped to two encoded binary bits $s_k^0$ and $s_k^1$ with respect to the generating rational function and the previous state. We use $L_k^b{\left( t \right)}$, $L_k^0{\left( t \right)}$, and $L_k^1{\left( t \right)}$ to denote the LLRs of $b_k$, $s_k^0$, and $s_k^1$ after the calculation of the BCJR algorithm for equivalent binary-input AWGN channel at the $t$-th iteration with mapping $s_k^0 = 0 \to {\bar s}_k^0 = {-1}$, $s_k^0 = 1 \to {\bar s}_k^0 = 1$, $s_k^1 = 0 \to {\bar s}_k^1 = {-1}$, and $s_k^1 = 1 \to {\bar s}_k^1 = 1$
\begin{equation}
{\bf y}_t = {\bf \bar s} + {\bf w}_t,
\end{equation}
where ${\bf w}_t$ is the real Gaussian noise with noise precision ${\eta}_x^t$. In addition, the corresponding conditional pdfs given $b_k$, $s_k^0$, and $s_k^1$ are represented by $p_{\left. {L_k^b}(t) \right| b_k}{( {{\cdot} | {\cdot}} )}$, $p_{\left. {L_k^0}(t) \right| {s_k^0}}{( {{\cdot} | {\cdot}} )}$, and $p_{\left. {L_k^1}(t) \right| {s_k^1}}{( {{\cdot} | {\cdot}} )}$, respectively. The conditional pdfs are identical for different encoded bits because the code length tends to infinity. Hence, we use $p_{\left. {L_t^b} \right| b}{\left( {{\cdot} | {\cdot}} \right)}$, $p_{\left. {L_t^0} \right| {s^0}}{\left( {{\cdot} | {\cdot}} \right)}$, and $p_{\left. {L_t^1} \right| {s^1}}{\left( {{\cdot} | {\cdot}} \right)}$ instead. Six conditional pdfs correspond to the LLRs are shown in Table II. The occurrence probabilities of encoded bits are given as $p{\left( s^0 \right)} = p$ and $p{\left( s^1 \right)} = 1 - p$. If the convolutional code is unbiased, then we have $p = {\frac{1}{2}}$. For unbiased convolutional code, the following equations hold approximately when code length tends to infinity:
\begin{subequations}
\begin{align}
p_{\left. {L_t^b} \right| b}{\left( {l|0} \right)} & = p_{\left. {L_t^b} \right| b}{\left( {-l|1} \right)}, \\
p_{\left. {L_t^0} \right| {s^0}}{\left( {l|0} \right)} & = p_{\left. {L_t^0} \right| {s^0}}{\left( {-l|1} \right)}, \\
p_{\left. {L_t^1} \right| {s^1}}{\left( {l|0} \right)} & = p_{\left. {L_t^1} \right| {s^1}}{\left( {-l|1} \right)}.
\end{align}
\end{subequations}
Fig.~\ref{pdf_1_2}(a) shows a case of the conditional pdfs. We find that the symmetry property of conditional pdfs holds for different constraint lengths. Since the encoded bits are mapped to the symbols uniformly and randomly under the uniform random interleaving, we have
\begin{align}
& {{\rm{mmse}_c}{\left( {\eta_x^t} \right)}} \nonumber\\
= & {\frac{1}{2}}{\int} {{\left( {1 - {\rm tanh}{\left( \frac{l}{2} \right)}} \right)}^2{\left( {p_{\left. {L_t^0} \right| {s^0}}{\left( {l|1} \right)}} + {p_{\left. {L_t^1} \right| {s^1}}{\left( {l|1} \right)}} \right)}dl}, \label{transfer_coded}
\end{align}
which can be obtained analogously as (\ref{transfer_uncoded}). However, the RHS of (\ref{transfer_coded}) cannot be obtained by numerical integration because the conditional pdfs ${p_{\left. {L_t^0} \right| {s^0}}{\left( {l|1} \right)}}$ and ${p_{\left. {L_t^1} \right| {s^1}}{\left( {l|1} \right)}}$ are intractable.

In the studies of iterative decoding, we can use the Gaussian pdfs to approximate such conditional pdfs, which is proved to be effective in the density evolution analysis when the code length tends to infinity \cite{mct}. Consider a binary-input AWGN channel with mapping $s = 0 \to x = {-1}$ and $s = 1 \to x = 1$
\begin{equation}\label{scalar_AWGN}
y_t = x + w_t,
\end{equation}
where $w_t$ is the real Gaussian noise with noise precision ${\eta}_{\rm s}^t$. The LLR of (\ref{scalar_AWGN}) is given as
\begin{equation}
L_s = {\ln}{\frac{p{\left( y|x = 1 \right)}}{p{\left( y|x = {-1} \right)}}} = {2{{\eta}_s^t}y}.
\end{equation}
Therefore, $L_s$ follows the Gaussian distribution ${\mathcal N}{\left( l; {-2{\eta}_s^t}, {4{\eta}_s^t} \right)}$ when $x = {-1}$ and ${\mathcal N}{\left( l; {2{\eta}_s^t}, {4{\eta}_s^t} \right)}$ when $x = 1$. We can use such Gaussian pdf to approximate the abovementioned conditional pdfs. The noise precision ${\eta}_s^t$ of Gaussian approximation can be determined by several methods, such as BER, SNR, and entropy matching \cite{mct}. We use the BER matching in this paper. In the BER matching, the noise precision ${\eta}_s^t$ is determined by matching the BER of Gaussian pdf and that of the conditional pdf
\begin{subequations}
\begin{align}
1 - {\Phi}{\left( {\sqrt {{\eta}_b^t}} \right)} & = {\int_{{-}{\infty}}^{0}}{p_{\left. {L_t^b} \right| b}{\left( {l|1} \right)}{\rm d}l}, \label{BCJR_b}\\
1 - {\Phi}{\left( {\sqrt {{\eta}_0^t}} \right)} & = {\int_{{-}{\infty}}^{0}}{p_{\left. {L_t^0} \right| {s^0}}{\left( {l|1} \right)}{\rm d}l}, \label{BCJR_s0}\\
1 - {\Phi}{\left( {\sqrt {{\eta}_1^t}} \right)} & = {\int_{{-}{\infty}}^{0}}{p_{\left. {L_t^1} \right| {s^1}}{\left( {l|1} \right)}{\rm d}l}, \label{BCJR_s1}
\end{align}
\end{subequations}
where the RHS of equations (\ref{BCJR_b}), (\ref{BCJR_s0}), and (\ref{BCJR_s1}) are estimated from the numerical simulations of the BCJR decoding. Then we have
\begin{subequations}
\begin{align}
{\eta}_b^t & = {\left[ {\Phi}^{-1}{\left( {1 - {\int_{{-}{\infty}}^{0}}{p_{\left. {L_t^b} \right| b}{\left( {l|1} \right)}{\rm d}l}} \right)} \right]^2}, \label{ber_match}\\
{\eta}_0^t & = {\left[ {\Phi}^{-1}{\left( {1 - {\int_{{-}{\infty}}^{0}}{p_{\left. {L_t^0} \right| {s^0}}{\left( {l|1} \right)}{\rm d}l}} \right)} \right]^2}, \label{bit0_match}\\
{\eta}_1^t & = {\left[ {\Phi}^{-1}{\left( {1 - {\int_{{-}{\infty}}^{0}}{p_{\left. {L_t^1} \right| {s^1}}{\left( {l|1} \right)}{\rm d}l}} \right)} \right]^2}. \label{bit1_match}
\end{align}
\end{subequations}
Fig.~\ref{pdf_1_2}(b) shows an instance of the conditional pdfs and its Gaussian approximations. Clearly, Gaussian pdfs are close to the corresponding conditional pdfs. Finally, according to the Gaussian approximation, we have
\begin{equation}\label{transfer_coded_app}
{{\rm{mmse}_c}{\left( {\eta_x^t} \right)}} = {\frac{1}{2}}{\left( {{\rm{mmse}_u}{\left( {\eta_0^t} \right)}} + {{\rm{mmse}_u}{\left( {\eta_1^t} \right)}} \right)},
\end{equation}
where ${\eta_0^t}$ and ${\eta_1^t}$ can be obtained from (\ref{bit0_match}) and (\ref{bit1_match}), respectively. Based on the above analysis, Proposition~1 can be obtained.

\end{appendices}

\end{document}